\documentclass[traditabstract, longauth]{aa}

\usepackage{graphicx}
\usepackage{txfonts}
\usepackage{natbib}
\usepackage{xspace}
\usepackage{hyperref}
\usepackage{scalerel}
\usepackage{multirow}
\usepackage[usenames]{xcolor}

\bibpunct{(}{)}{;}{a}{}{,}

\newcommand{\pycs}{{\tt PyCS}\xspace}

\newcommand{\hc}{$H_0$}
\def\ksmpc{${\rm\, km\, s^{-1}\, Mpc^{-1}}$\xspace}

\newcommand{\bd}{\begin{displaymath}}
\newcommand{\ed}{\end{displaymath}}
\newcommand{\be}{\begin{equation}}
\newcommand{\ee}{\end{equation}}
\newcommand{\beaa}{\begin{eqnarray*}}
\newcommand{\eeaa}{\end{eqnarray*}}
\newcommand{\bea}{\begin{eqnarray}}
\newcommand{\eea}{\end{eqnarray}}

\def\hequad{HE\,0435$-$1223}

\def\wfilens{WFI2033$-$4723}
\def\blens{B1608$+$656}
\def\rxjlens{RXJ1131$-$1231}

\def\kmsMpc {\rm km\,s^{-1}\,Mpc^{-1}}

\definecolor{darkgreen}{rgb}{0.0, 0.5, 0.0}

\definecolor{crimson}{HTML}{DC143C}
\definecolor{royalblue}{HTML}{4169E1}
\definecolor{hotpink}{HTML}{FF69B4}
\definecolor{purple}{HTML}{800080}
\definecolor{midnightblue}{HTML}{191970}

\definecolor{indianred}{HTML}{CD5C5C}
\definecolor{seagreen}{HTML}{2E8B57}
\definecolor{steelblue}{HTML}{4682B4}
\definecolor{darkorange}{HTML}{FF8C00}
\definecolor{brown}{HTML}{A52A2A}

\def\pycs{{\tt PyCS}}
\def\hc{$H_0$\xspace}
\def\LCDM{$\Lambda {\rm CDM}$\xspace}

\begin{document}

\title{COSMOGRAIL XVIII: time delays of the quadruply lensed quasar \wfilens.}
\titlerunning{Time-delay measurements of \wfilens}

\author{
V.~Bonvin\inst{\ref{epfl}} \and
M.~Millon\inst{\ref{epfl}} \and
J.~H-H.~Chan\inst{\ref{epfl}} \and
F.~Courbin\inst{\ref{epfl}} \and
C.~E.~Rusu\inst{\ref{UCD}, \ref{Subaru}} \and
D.~Sluse\inst{\ref{STAR}} \and
S.~H.~Suyu\inst{\ref{MPG}, \ref{TUM}, \ref{AASIA}} \and
K.~C.~Wong\inst{\ref{IPMU}, \ref{NAOJ}} \and
C.~D.~Fassnacht\inst{\ref{UCD}} \and
P.~J.~Marshall\inst{\ref{KIPAC}} \and
T.~Treu\inst{\ref{UCLA}} \and
E.~Buckley-Geer \inst{\ref{fermilab}} \and
J.~Frieman \inst{\ref{fermilab}, \ref{chicago}} \and
A.~Hempel \inst{\ref{UNAB}, \ref{heidelberg}} \and 
S.~Kim \inst{\ref{puc}, \ref{heidelberg}}\and
R.~Lachaume \inst{\ref{puc}, \ref{heidelberg}}\and 
M.~Rabus \inst{\ref{puc}, \ref{heidelberg}, \ref{LCOGT}, \ref{UCSB}}\and
D.~C.-Y.~Chao \inst{\ref{MPG}, \ref{TUM}} \and
M.~Chijani \inst{\ref{UNAB}} \and
D.~Gilman \inst{\ref{UCLA}} \and
K.~Gilmore \inst{\ref{KIPAC}} \and
K.~Rojas \inst{\ref{epfl}} \and
P.~Williams \inst{\ref{UCLA}} \and
T.~Anguita \inst{\ref{UNAB}, \ref{millenium}} \and
C.~S.~Kochanek \inst{\ref{AOSU}, \ref{CCAPOSU}} \and
G.~Meylan \inst{\ref{epfl}} \and
C.~Morgan \inst{\ref{USNA}} \and
V.~Motta \inst{\ref{valpo}} \and
M.~Tewes\inst{\ref{bonn}}
}

\institute{
Institute of Physics, Laboratory of Astrophysics, Ecole Polytechnique 
F\'ed\'erale de Lausanne (EPFL), Observatoire de Sauverny, 1290 Versoix, 
Switzerland \label{epfl}\goodbreak \and
Max Planck Institute for Astrophysics, Karl-Schwarzschild-Str.1, D-85740 Garching, Germany \label{MPG}\goodbreak \and
Physik-Department, Technische Universit\"at M\"unchen, 
James-Franck-Stra\ss{}e~1, 85748 Garching, Germany \label{TUM} \goodbreak \and
Institute of Astronomy and Astrophysics, Academia Sinica, 11F of ASMAB, No.1, Section 4, Roosevelt Road, Taipei 10617, Taiwan \label{AASIA} \goodbreak \and
Subaru Telescope, National Astronomical Observatory of Japan, 650 N Aohoku Pl, Hilo, HI 96720 \label{Subaru} \goodbreak \and
Department of Physics, University of California, Davis, CA 95616, USA \label{UCD} \goodbreak \and
STAR Institute, Quartier Agora - All\'ee du six Ao\^ut, 19c B-4000 Li\`ege, Belgium \label{STAR} \goodbreak \and
Kavli IPMU (WPI), UTIAS, The University of Tokyo, Kashiwa, Chiba 277-8583, Japan \label{IPMU} \goodbreak \and
National Astronomical Observatory of Japan, 2-21-1 Osawa, Mitaka, Tokyo 181-8588, Japan \label{NAOJ} \goodbreak \and
Department of Physics and Astronomy, PAB, 430 Portola Plaza, Box951547, Los Angeles, CA 90095, USA \label{UCLA} \goodbreak \and
Kavli Institute for Particle Astrophysics and Cosmology, Stanford University, 452 Lomita Mall, Stanford, CA 94035, USA \label{KIPAC} \goodbreak \and
Fermi National Accelerator Laboratory, P.O. Box 500, Batavia, IL 60510, USA \label{fermilab}\goodbreak \and
Kavli Institute for Cosmological Physics, University of Chicago, Chicago, IL 60637, USA \label{chicago}\goodbreak \and
Departamento de Ciencias F\'isicas, Universidad Andres Bello Fernandez 
Concha 700, Las Condes, Santiago, Chile  \label{UNAB}\goodbreak \and
Centro de Astroingenier\'ia, Facultad de F\'isica, Pontificia Universidad 
Cat\'olica de Chile, Av. Vicu\~na Mackenna 4860, Macul 7820436, 
Santiago, Chile \label{puc}\goodbreak \and
Max-Planck-Institut f\"ur Astronomie, K\"onigstuhl 17, 69117 Heidelberg, 
Germany \label{heidelberg}\goodbreak \and
Las Cumbres Observatory Global Telescope, 6740 Cortona Dr., Suite 102, Goleta, CA 93111, USA \label{LCOGT}\goodbreak \and
Department of Physics, University of California, Santa Barbara, CA 93106-9530, USA \label{UCSB}\goodbreak \and
Instituto de F\'isica y Astronom\'ia, Universidad de Valpara\'iso, Avda. 
Gran Breta\~na 1111, Playa Ancha, Valpara\'iso 2360102, Chile 
\label{valpo}\goodbreak \and
Millennium Institute of Astrophysics, Chile \label{millenium}\goodbreak \and
Department of Physics, United States Naval Academy, 572C Holloway Rd., Annapolis, MD 21402, USA \label{USNA}\goodbreak \and
Argelander-Institut f\"ur Astronomie, Auf dem H\"ugel 71, 53121, Bonn, 
Germany \label{bonn}\goodbreak \and
Department of Astronomy, The Ohio State University, 140 West 18th Avenue, Columbus, OH 43210, USA \label{AOSU}\goodbreak \and
Center for Cosmology and Astroparticle Physics, The Ohio State University, 191 W. Woodruff Avenue, Columbus, OH 43210, USA \label{CCAPOSU}
}

\date{\today}
\abstract{We present new measurements of the time delays of \wfilens. The data sets used in this work include 14 years of data taken at the 1.2m Leonhard Euler Swiss telescope, 13 years of data from the SMARTS 1.3m telescope at Las Campanas Observatory and a single year of high-cadence and high-precision monitoring at the MPIA 2.2m telescope. The time delays measured from these different data sets, all taken in the $R$-band, are in good agreement with each other and with previous measurements from the literature. Combining all the time-delay estimates from our data sets results in $\Delta t_{AB} = 36.2_{-0.8}^{+0.7}$ days (2.1\% precision),  $\Delta t_{AC} = -23.3_{-1.4}^{+1.2}$ days (5.6\%) and $\Delta t_{BC} = -59.4_{-1.3}^{+1.3}$ days (2.2\%). In addition, the close image pair A1-A2 of the lensed quasars can be resolved in the MPIA 2.2m data. We measure a time delay consistent with zero in this pair of images. We also explore the prior distributions of microlensing time-delay potentially affecting the cosmological time-delay measurements of \wfilens. There is however no strong indication in our measurements that microlensing time delay is neither present nor absent. This work is part of a H0LiCOW series focusing on measuring the Hubble constant from \wfilens.}


\maketitle

\section{Introduction}

The flat-\LCDM model, often labelled Standard Cosmological Model due to its ability to fit extremely well most of today's cosmological observations, has recently been strengthened by the final update of the Planck satellite CMB observations \citep{Planck2018overview, Planck2018cosmo}. However, a few sources of tension remain. One example is the predicted amplitude of overdensities in the Universe, characterized by the normalisation of the linear matter power spectrum $\sigma_8$, that is in mild tension with direct measurements, such as the cosmic-shear analyses from KiDS \citep{Kohlinger2017, Troxel2018} and HSC \citep{Hikage2019}. Similarly, Lyman-$\alpha$ measurements of Baryon Acoustic Oscillations at intermediate redshift \citep[e.g.][]{Bautista2017} are in a mild tension with the flat-\LCDM\ predictions. More stringent is the measured expansion rate of the Universe, also called the Hubble constant \hc, constrained through the observation of standard candles \citep[e.g.][]{Riess2019} that is in $4.4 \sigma$ tension with the flat-\LCDM\ predictions.

Whether such tensions result from underestimated systematic errors in at least one of the measurements, a statistical fluke, or point towards new physics beyond the Standard Model is currently a topic of discussion \citep[see e.g.][for recent examples]{Mortsell2018, Poulin2018, Amendola2019, Capozziello2019, Pandey2019}. There is however no simple, preferred alternative or extension to the flat-\LCDM\ model that is clearly favoured by the data at the moment. In such a context, the way forward consists of improving the precision and accuracy of all measurements involved, as well as using alternative and independent techniques to estimate the conflicting cosmological parameters. Focusing on the Hubble constant, the so-called distance ladder method based on the cross-calibration of various distance indicators offers multiple routes towards \hc\ \citep[see e.g.][for recent updates]{Cao2017, Jang2018, Dhawan2018, Riess2019}. Galaxy clustering offers an alternative way to measure the Hubble constant independently from CMB measurements \citep[e.g.][]{DES2018, Kozmanyan2019}, as well as the so-called ``standard sirens'' technique based on gravitational waves events \citep[e.g.][]{Abbott2017, ChenHY2018, Feeney2019} or the observations of water vapor megamasers \citep[e.g.][]{Reid2013, Braatz2018}.

An independent approach to directly measure \hc\ is to use time-delay cosmography. The idea, first proposed by \citet{Refsdal1964}, consists of measuring the time delay(s) between the luminosity variations of multiple images of a strongly gravitationally lensed source. Combined with careful modeling of the mass distribution of the lens galaxy, its surroundings and accounting of the mass along the line-of-sight, such measurement provide a direct way to constrain \hc, which is nearly independently of any other cosmological parameters \citep[see][for a review]{Treu2016, Suyu2018}.

The H0LiCOW collaboration \citep{Suyu2017} focuses on this method, using state-of-the-art softwares and techniques to carry out each step of the analysis. In 2017, H0LiCOW released its first results based on a blind analysis of the quadruply lensed quasar \hequad\ \citep{Sluse2017, Rusu2017, Wong2017, Bonvin2017, Tihhonova2018}. Combined with two other strongly lensed systems analysed earlier \citep{Suyu2010, Suyu2014} and a fourth system analyzed recently \citep{Birrer2019}, it resulted in a 3\% precision determination of the Hubble constant in a flat-\LCDM universe, \hc$=72.5_{-2.3}^{+2.1}\kmsMpc$. This result, in mild tension with the CMB predictions but in excellent accordance with the distance-ladder measurements, proves both the robustness of the method and its potential for deciding whether the discrepancies seen in $H_0$ measurements are real or not.

The present paper is part of a series focusing on the analysis of the quadruply lensed quasar \wfilens. It presents new time delay measurements based on monitoring data from the COSMOGRAIL collaboration taken between 2004 and 2018. In parallel, Sluse et al. 2019 (submitted; hereafter H0LiCOW XI) presents measurements of the spectroscopic redshifts of galaxies in the environment of \wfilens. Rusu et al. 2019, submitted (hereafter H0LiCOW XII) focuses on the modeling of \wfilens, taking into account the environment from H0LiCOW XI and the time delays from the present work, to derive a value of the Hubble constant. Finally, Wong et al. 2019, in prep. (hereafter H0LiCOW XIII) combines the time-delay distances from all the strongly lensed systems analysed so far by the H0LiCOW collaboration to infer cosmological parameters. 

The present manuscript is divided as follows: Section~\ref{sec:data} presents the monitoring campaigns and the data reduction process that yield the light curves of the lensed images of the quasar. Section~\ref{sec:timedelays} presents the time-delay measurement framework, its application to our light curves and a series of robustness tests. Section~\ref{sec:mltd} quantifies the effect of the microlensing time delay on the time-delay measurements. Section~\ref{sec:conclusions} summarizes our results and gives our conclusions.

\section{Data sets}

\label{sec:data}

\begin{table*}
 \caption{Summary of the optical monitoring campaigns of \wfilens. Sampling refers to the targeted cadence, not counting the seasonal gaps. The number of epochs ($\#$obs), median seeing, airmass and total exposure time are computed from the epochs used in the light curves presented in Fig.~\ref{fig:lcs}, which discard $\sim 13\%$ of the total number of exposures for each data set (mostly due to bad weather or scheduling conflicts). The distribution in seeing and airmass of each data set is shown in the bottom-right panel of Fig.~\ref{fig:lcs}.}
  \centering
 \begin{tabular}{l c c r r r c c r}
  \hline
  Telescope-Instrument & FoV & Pixel & Period of observation & $\#$obs & 
Exp.time & Seeing & Airmass & Sampling\\
  \hline
  Euler-C2 & 11'$\times$11' & 0.344'' & Oct 2004 - Sep 2010 & 294 & 
5$\times$360s &  1.59'' & 1.18& 6 days \\
  Euler-ECAM & 14.2'$\times$14.2' & 0.215'' & Oct 2010 - May 2018 & 
350 & 5$\times$360s &  1.54''&1.15 & 4 days \\
  SMARTS-ANDICAM & 10'$\times$10' & 0.300'' & Apr 2004 - Nov 2016 & 
345 & 3$\times$300s & 1.50'' & 1.17& 4 days \\
  2.2m-WFI & 36'$\times$36' & 0.238'' & Mar 2017 - Dec 2017 & 
136 & 4$\times$320s & 1.34'' & 1.17&1 day \\

  \hline
  {\bf TOTAL} & - & - & Oct 2004 - May 2018 & 876 & 447.0h & - & - \\
  \hline
 \end{tabular}
 \label{tab:obslog}
\end{table*}

\begin{figure*}
  \centering
  \includegraphics[width=0.99\textwidth]{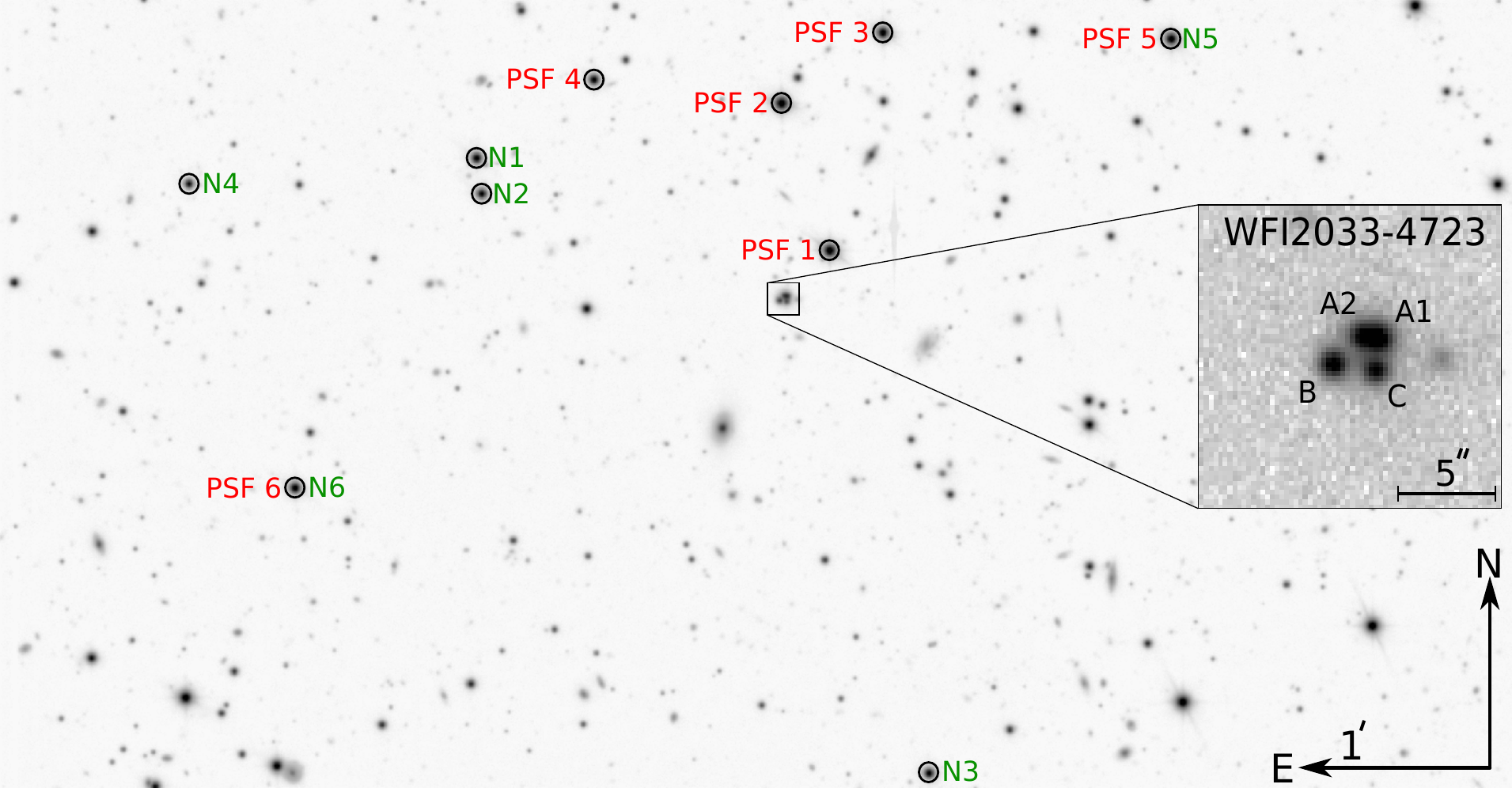}
  \caption{Part of the field of view of \wfilens\, seen through the WFI instrument at the 2.2m MPIA telescope. The field is a stack of 312 images with seeing $\leq 1.4''$ and ellipticity $\leq 0.18$, totalling $\sim 28$ hours of exposure. The insert represent a single, 320s exposure of the lens in $0.65''$ seeing. The stars labelled PSF~1 to PSF~6 in red were used to build the point spread function. The stars labelled N1 to N6 in green were used to compute the exposure-to-exposure normalisation coefficient.}
  \label{fig:fov}
\end{figure*}

\wfilens\ is a bright, quadruply lensed quasar in a fold configuration \citep[$\alpha$(2000):~20h33m42.08s; $\delta$(2000):~-47$^\circ$23'43.0",][]{Morgan2004}. The most recent spectroscopic measurements of the lens and source redshifts are $z_d=0.6575$ (H0LiCOW XI) and $z_s=1.662$ \citep{Sluse2012}, respectively.

\wfilens\ has been monitored since 2004 by the 1.2m Swiss Leonhard Euler telescope at the ESO La Silla Observatory in Chile, using the C2 instrument until October 2010 and the ECAM instrument since then, using the {\tt Rouge Gen\`eve} filter, a modified version of the broad {\tt Bessel-R} filter. Parallel observations took place since 2004 with the 1.3m Small and Moderate Aperture Research System (SMARTS) at the Cerro Tololo Inter-American Observatory (CTIO), using an {\tt KPNO R-band} filter. From March to December 2017, \wfilens\ was also monitored daily at the MPIA 2.2m telescope at the ESO La Silla Observatory using the WFI instrument, using the {\tt ESO BB\#Rc/162} filter. A summary of the observations is presented in Table \ref{tab:obslog}. The main steps of the data reduction process are as follows:

\begin{enumerate}

\item Bias and flat exposures are taken on a regular basis. Master biases exposures are constructed to remove the additional bias level, and master flats exposures are constructed to correct from the difference of the pixel sensitivity of the CCD detector. On the WFI exposures, fringes are occasionally observed. A fringe pattern correction is constructed by sigma clipping and stacking dithered exposures from the same epoch. The pattern is then subtracted from each individual science exposure.\\

\item The cleaned science exposures are sky-subtracted and aligned using standard {\tt sextractor} and {\tt IRAF} procedures. A number of reference stars are selected to construct an empirical Point Spread Function (PSF). Figure \ref{fig:fov} presents a stacked exposure of the field-of-view of \wfilens\ from the WFI instrument. The stars selected for the PSF construction are circled in green and labeled PSF 1 to PSF 6. Various choices of reference stars were explored, with the best results (i.e. the smallest residuals after deconvolution) being obtained for stars close in projection to the lens and slightly brighter than the lensed quasar images.\\

\item The quasar images are deconvolved, using the empirical PSF and the MCS deconvolution algorithm \citep{Magain1998, Cantale2016}. The deconvolution process separates the quasar light into two channels: a) an analytical channel representing the quasar images as point sources, variable from exposure to exposure, and b) a numerical channel representing the light from the lens galaxy, other extended sources (such as the host galaxy of the quasar) as well as nearby perturbers. The flux of the point sources representing the quasar images is then normalised by an exposure-to-exposure normalisation coefficient, which is computed using a selection of reference stars in the field (circled in green and labelled N1 to N6 on Fig.~\ref{fig:fov}) whose flux has been estimated in the same manner as the quasar images. This allows to minimize the effect of systematic errors due to the deconvolution process and PSF mismatches. The normalised fluxes of each lensed quasar image are then combined for each observing epoch.\\

\end{enumerate}

\begin{figure*}
  \centering
  \includegraphics[width=0.99\textwidth]{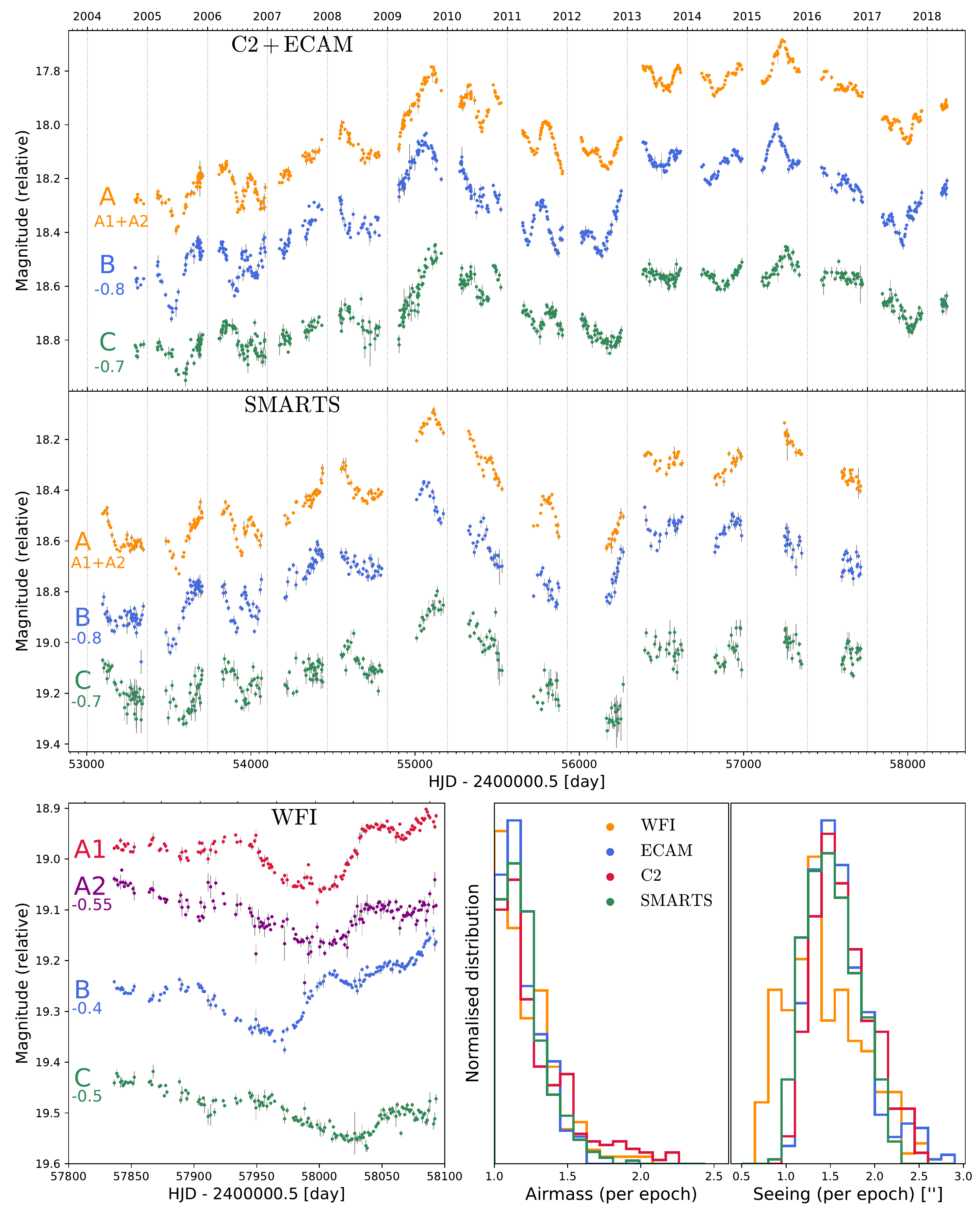}
  \caption{\wfilens\ light curves from the four different instruments used in this work. The top panel presents the Euler data sets (C2+ECAM instruments), with the change occuring in October 2010, corresponding to the small gap visible in the 2010 season. The middle panel presents the SMARTS data set. The bottom left panel presents the WFI data set, where the A1 and A2 images were individually resolved thanks to superior image quality and longer exposure times. We note that different calibration stars were used for the different data sets, hence the different relative magnitudes values between different instruments. The bottom right panel presents the normalised distribution of the airmass and seeing of all the individual exposures in each data set.}
  \label{fig:lcs}
\end{figure*}

The light curves obtained from each instrument are presented in Fig.~\ref{fig:lcs}. The top panel presents the combined C2+ECAM data sets (the small gap in the 2010 season corresponding to the change of instrument), the middle panel presents the SMARTS data set and the bottom-left panel presents the WFI data set. Similar features are clearly visible across all light curves and data sets. A merger of all the data sets exhibits no discrepancies in the stacked light curves, provided an instrumental offset in magnitude between each data set. We note that the C2, SMARTS and ECAM data sets have been already partially discussed: \citet{Vuissoz2008} use three and a half years of data (from 2004 to mid-2007) to measure time delays and the Hubble constant with crude models, and \citet{Morgan2018} use the SMARTS, C2 and ECAM data up to 2016 to estimate the quasar accretion disk size from microlensing analysis. We emphasize that the data reduction process presented in this section and the time-delay analysis presented in Sec.~\ref{sec:timedelays} are completely independent from previous studies. The ECAM data set after 2016 and the entire WFI data set are presented for the first time in this work. The ambient quality of the observations are presented in the bottom right panel of Fig.~\ref{fig:lcs}. The WFI data set clearly stands out as the best in seeing, thanks to better dome seeing, better instrument and flexible scheduling at the telescope.

For the C2, ECAM and SMARTS data sets, our deconvolution scheme is not able to properly resolve the close pair of images A1 and A2. The two deconvolved images share flux even after deconvolution, and the structures in their light curves are clearly polluted by correlated noise. As the cosmological time delay between the A1 and A2 images is expected to be very small \citep{Vuissoz2008}, we chose instead to merge A1 and A2 into a single, virtual image A light curve. For the WFI data set, the size of the instrument's primary mirror (2.2 metres) coupled to a longer exposure time with respect to the other instruments allows us to reach a sufficient signal-to-noise ratio to properly resolve the A1 and A2 images after deconvolution. An estimate of the time delay between A1 and A2 is given in Sec.~\ref{sec:timedelays}. Having an extra light curve allows a third independent time-delay measurement to constrain the lens models. In addition, it solves the issue of the anchoring position of a measured delay using a joint light curve that arises at the lens modeling stage. As such, it represents a crucial extra piece of information for time-delay cosmography.

\section{Time-delay measurement}

\label{sec:timedelays}

In this section, we measure the time delays for each of the individual data sets and combine them into a final time-delay estimate to be used at the lens modeling stage (see H0LiCOW XII). We start by giving a description of the framework used in this paper, emphasizing how the time-delay uncertainties are evaluated, before presenting our time-delay estimates and performing robustness checks to assess the quality of our measurements. Let us first recall the terminology used here, adopted from \citet{Tewes2013a, Bonvin2018, Bonvin2019}.

\begin{itemize}

 \item A \emph{point estimator} is a method that returns the best time-delay estimates $\mathbf{\Delta t} = [ \Delta t_{ij} ]$ between two or more light curves, with ${i, j} \in [A, B, C,...]$. A point estimator depends on \emph{estimator parameters} that impact its results, for example by controlling the smoothness of a fit.
 
 \item The \emph{intrinsic error} is the dispersion of a point estimator when randomizing its initial state, such as the starting point of an iterative estimator (in our case a guess time delay). We note that this is different from the error we would get by varying the estimator parameters.
 
 \item A \emph{generative model} of mock light curves is a process that draws simulated light curves that mimick as closely as possible the input data, but with known time delays. We apply our point estimators on a large set of these simulated light curves in order to estimate the point estimators uncertainties $\mathbf{\delta t} = [ (\delta t_{+}, \delta t_{-})_{ij} ]$. Generative models, similarly to point estimators, depend on a set of \emph{generative model parameters} that control the generation of the simulated light curves.
 
 \item A \emph{curve-shifting technique} regroups a point estimator with a chosen set of estimator parameters in order to estimate the time delays from the data, and a generative model with a fixed set of parameters to draw mock light curves with known time delays from which the uncertainty of the point estimator is assessed. In the following, we call the \emph{curve-shifting technique parameters} the joint point estimator and generative model parameters.
 
 \item A \emph{Group} $G=[E_{ij}]$, composed of independent \emph{time-delay estimates $E = \Delta t^{+\delta t_{+}} _{-\delta t_{-}}$} is the result of the application of a curve-shifting technique on a data set.
 
 \item A \emph{Series} $S=[G_1,... G_k, ..., G_N], k \in N$, consists of multiple Groups that share the same data set and point estimator, but differ in their choice of point estimator parameters and/or generative model parameters. Typically, a Series is a collection of Groups whose time-delay estimates are \emph{a priori} equivalently probable, and thus can be combined or marginalised over.
 
\end{itemize}

\begin{figure*}
  \centering
  \includegraphics[width=0.99\textwidth]{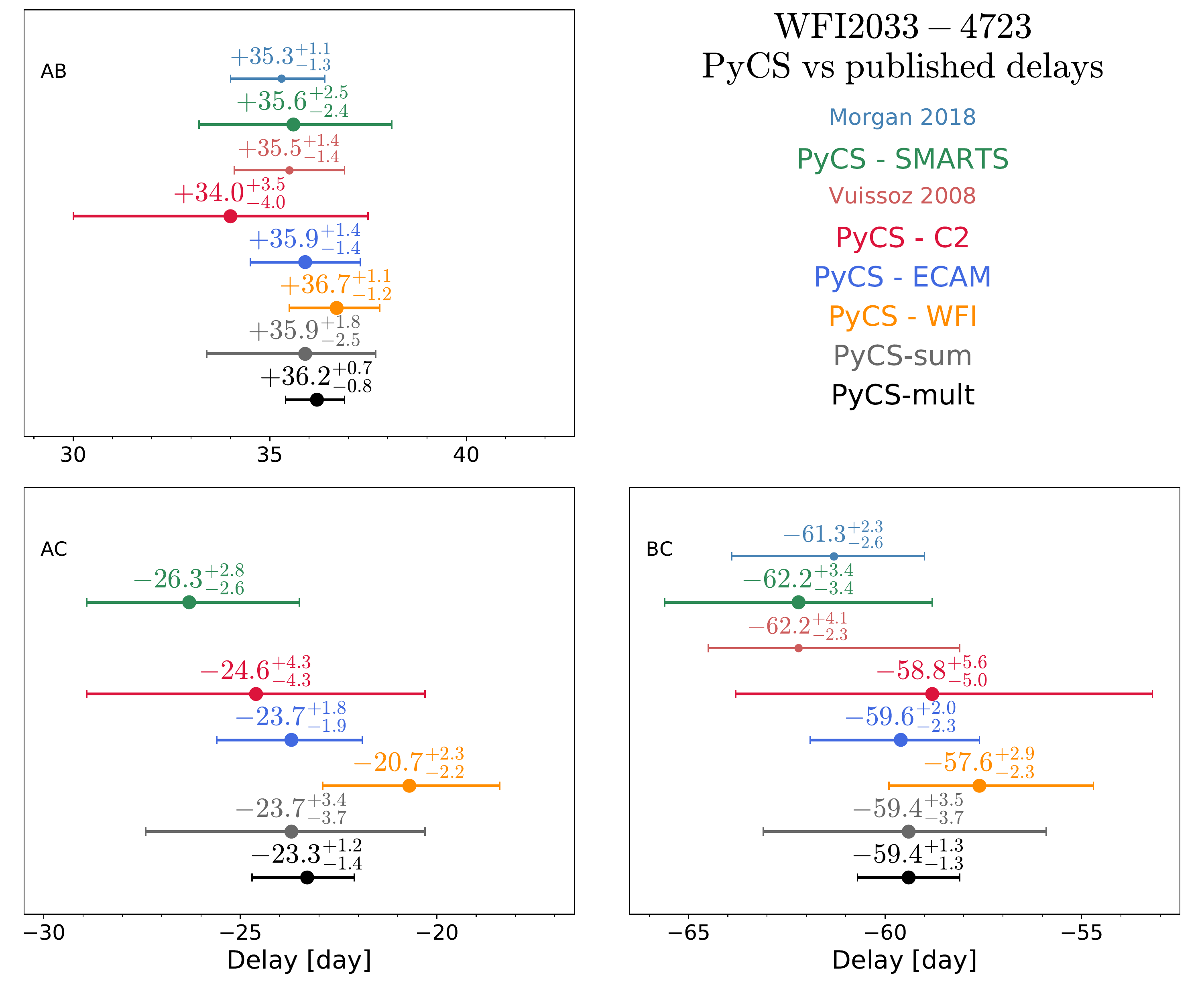}
  \caption{\wfilens\ time-delay estimates. The colored points labelled ``PyCS'' represent the final time-delay estimates for each data sets, obtained after combining  the curve-shifting technique parameters following the marginalisation scheme presented in Section \ref{sec:pycs}. The gray and black points, respectively labelled ``PyCS-sum'' and ``PyCS-mult'' represent the marginalisation and combination of the results obtained on the individual data sets. Indicated with smaller points are the AB and BC delays taken from \citet{Morgan2018} and \citet{Vuissoz2008}, respectively. The values indicated above each measurement represent the $50^{\mathrm{th}}$, $16^{\mathrm{th}}$ and $84^{\mathrm{th}}$ percentiles of the respective probability distributions.}
  \label{fig:delays}
\end{figure*}


\begin{table*}
\centering
  \caption{List of the estimator parameters used to compute the time-delay estimates presented in Fig.~\ref{fig:delays}. $\eta$ corresponds to the initial knot spacing of the intrinsic spline and $\eta_{\mathrm{ml}}$ to the initial knot spacing of the extrinsic microlensing splines. For the WFI data set, ``ml model'' indicates whether the microlensing has been modeled using free-knot splines or polynomials. For the former case, $\eta_{\mathrm{ml}}$ pos. indicates the constraint on the microlensing spline knots. Brackets indicate that the values within have been tested in all possible combinations - each data set has thus nine different possible combinations. For the regression difference technique the parameters $\nu$ (smoothness degree), A (amplitude in magnitudes), scale (length scale in days) and errscale (observation variance in days) refer to the Mat\'ern covariance function used in the \texttt{python 2.7} Gaussian process regression implementation of the \texttt{pymc.gp} module. The rightmost column (marked with an *) has the Mat\'ern covariance function replaced by a power-law covariance function, where $\nu$ indicates the power-law index.}

  \begin{tabular}{l | c c c c | c c c c c c}
    \multicolumn{1}{c}{} & \multicolumn{4}{c|}{free-knot splines} & \multicolumn{5}{|c}{regression difference} & (*) \\ \hline\hline

    \multirow{4}{*}{C2} & \multirow{2}{*}{$\eta$} & \multicolumn{3}{c|}{\multirow{2}{*}{[25, 35, 45]}} &
    $\nu$ & 1.7 & 2.2 & 1.9 & 1.3 & 1.8 \\
    
    & & \multicolumn{3}{c|}{} & A & 0.5 & 0.4 & 0.6 & 0.3 & 0.7 \\
    
    & \multirow{2}{*}{$\eta_{\mathrm{ml}}$} & \multicolumn{3}{c|}{\multirow{2}{*}{[150, 300, 600]}} & 
    scale & 200 & 200 & 200 & 300 & 250 \\    
    
    & & \multicolumn{3}{c|}{} & errscale & 20 & 25 & 20 & 25 & 25  \\ \hline

    \multirow{4}{*}{ECAM} & \multirow{2}{*}{$\eta$} & \multicolumn{3}{c|}{\multirow{2}{*}{[25, 35, 45]}} &
    $\nu$ & 1.7 & 2.2 & 1.5 & 1.3 & 1.8 \\
    
    & & \multicolumn{3}{c|}{} & A & 0.5 & 0.4 & 0.4 & 0.2 & 0.3 \\
    
    & \multirow{2}{*}{$\eta_{\mathrm{ml}}$} & \multicolumn{3}{c|}{\multirow{2}{*}{[150, 300, 600]}} & 
    scale & 200 & 200 & 200 & 200 & 200 \\    
    
    & & \multicolumn{3}{c|}{} & errscale & 20 & 25 & 20 & 5 & 5  \\ \hline

    \multirow{4}{*}{SMARTS} & \multirow{2}{*}{$\eta$} & \multicolumn{3}{c|}{\multirow{2}{*}{[35, 45, 55]}} &
    $\nu$ & 2.2 & 1.8 & 1.9 & 1.3 & 1.7 \\
    
    & & \multicolumn{3}{c|}{} & A & 0.5 & 0.7 & 0.6 & 0.3 & 0.7 \\
    
    & \multirow{2}{*}{$\eta_{\mathrm{ml}}$} & \multicolumn{3}{c|}{\multirow{2}{*}{[150, 300, 600]}} & 
    scale & 200 & 200 & 200 & 150 & 300 \\    
    
    & & \multicolumn{3}{c|}{} & errscale & 20 & 25 & 20 & 10 & 25  \\ \hline

    \multirow{4}{*}{WFI} & $\eta$ & \multicolumn{3}{c|}{[25, 35, 45]} &
    $\nu$ & 1.7 & 1.8 & 1.3 & 1.5 & 1.9 \\

    & \multirow{2}{*}{ml model} & \multicolumn{2}{c|}{\multirow{2}{*}{splines}} & \multirow{2}{*}{$3^{\mathrm{rd}}$ poly.} & 
    A & 0.5 & 0.6 & 0.3 & 0.4 & 0.7 \\
    
    & & \multicolumn{3}{c|}{} & scale & 200 & 200 & 200 & 150 & 300 \\    
    
    & $\eta_{\mathrm{ml}}$ pos. & \multicolumn{2}{c|}{[free, fixed]} & - & 
    errscale & 20 & 25 & 20 & 10 & 25  \\ \hline

  \end{tabular}

\label{tab:tabparams}  
\end{table*}

\subsection{PyCS formalism}
\label{sec:pycs}

We measure the time delays using the open-source \pycs\ software \citep{Tewes2013a, Bonvin2016}. \pycs\ is currently composed of two different point estimators.

\begin{itemize}
 \item The \emph{free-knot splines} estimator fits i) a unique spline to model the intrinsic luminosity variations common to all the lensed images, and ii) individual extrinsic splines to each light curve in order to model the microlensing magnification by compact objects in the lens galaxy. The time shifts between the light curves are optimised concurrently with the splines fits in an iterative process, following the BOK-splines algorithm of \citet{Molinari2004}. The smoothness of the fit is controlled by the number of knots in the splines, which we represent by the initial (i.e. prior to the optimisation) temporal spacing between the knots, denoted $\eta$ and $\eta_{ml}$ for the intrinsic and extrinsic splines, respectively. Special constraints on the knots position are denoted by $\eta^{pos}$ and $\eta_{ml}^{pos}$.
 
 \item The \emph{regression difference} estimator fits individual regressions through each light curves using Gaussian Processes. The regressions are then shifted in time with respect to each other, pair by pair. At each time step, the amount of variations of the difference of the two regressions is evaluated; to the smallest variability (the ``flatest'' difference curve) is associated the best time-delay point estimate. The estimator parameters consist of a choice of covariance kernel (and associated parameters) for the Gaussian Process. The regression difference estimator does not require any explicit modeling of the microlensing. 
\end{itemize}

To assess the uncertainties of these two point estimators, \pycs\ implements a generative model for mock lensed quasar light curves. It takes as input i) a model for the intrinsic luminosity variations of the quasar, ii) models for each individual extrinsic variations of the light curves, iii) the sampling and noise properties of the observations, and iv) the true time delays to recover. The intrinsic and extrinsic variations are modeled after the best fit of the free-knot splines estimator on the real data. The sampling of the mock curves matches the one of the real observations, and the noise is a combination of red and white noise that are calibrated on the real data \citep[see][for the details]{Tewes2013a}. Finally, the true time delays are randomly sampled over a range of values around the actual time delays measured on the real light curves. The dispersions of the point estimators' results around the true delays of the mock light curves are used to estimate the uncertainties.

With two point estimators and one generative model, \pycs\ offers two different curve-shifting techniques. The two techniques are not completely independent as they share the same generative model, but offer a nice cross-validation of the results. Note that we are exploring a novel generative model in our robustness tests, as compared with previous work. When having access to multiple data sets from different instruments, like in the present case for \wfilens, we chose to analyse the data sets independently from each other as long as they are each of sufficient quality to yield robust time-delay estimates. Proceeding this way is also better to fully extract the constraining power of each data set, since different data sets can be best fit with different estimator parameters due to their differences in sampling and photometric uncertainties. We explore the results obtained when combining the data sets as a robustness test in Section \ref{sec:robustness}.

For each curve-shifting technique and each data set, we vary the curve-shifting technique parameters to obtain a Group of time-delay estimates. Exploring the range of plausible curve-shifting parameters is done as follows. For a curve-shifting technique that uses the free-knot splines estimator, we use the same parameters for the point estimator and the generative model (recall that the generative model is based on a free-knot spline fit to the data), for simplicity. For a curve-shifting technique that uses the regression difference estimator, we use the same generative model parameters as the curve-shifting techniques based on the free-knot splines estimator. What constrains our choice of parameters for a point estimator is the intrinsic error it yields; we limit ourselves to the sets of parameters that yield an intrinsic error qualitatively much smaller than the smallest total uncertainty of the most precise curve-shifting technique we have. Although this might seem somewhat arbitrary, a large intrinsic error is a robust indicator that the point estimator either overfit or underfit the data, which is something we want to avoid.

The various Groups obtained by exploring the different plausible sets of curve-shifting technique parameters are then regrouped in a Series. Each Group of the Series represents a plausible measurement of the time delays. If all the Groups have similar time-delay measurements (albeit with different uncertainties), we consider the most precise Group as our reference. However, if two or more Groups are in tension, keeping only the most precise one is not a good option as the tension might indicate a possible bias related to the choice of the curve-shifting parameters. Instead, we iteratively marginalise over the Groups in tension until that tension stays below a defined threshold, following the formalism presented in Section 4.1 of \citet{Bonvin2019}. We adopt a fiducial threshold value of $\sigma_{\rm{thresh}}=0.5$ in this work. In practice, it means that once the most precise Group of the Series has been identified, its tensions with the remaining Groups are computed; the most precise Group for which the tension with the reference Group exceeds 0.5$\sigma$ (if any) is selected, and the marginalisation (summation) of these two Groups become the new reference Group. The process is repeated until all the tensions drop below 0.5$\sigma$, or all the Groups are combined (which, in the present case, never happens). As the choice of $\sigma_{\rm{thresh}}=0.5$ might seem arbitrary, we explore the impact of varying this threshold in our robustness tests in Section \ref{sec:robustness}. 

In first approach, exploring the curve-shifting technique parameters and combining the results as we do in this work can be done using a grid search followed by a weighted marginalisation over the results. Ideally, the exploration of the curve-shifting parameters would be performed in a full Bayesian framework, but this is currently not achievable computationally due to the time required to sample the results for a single set of curve-shifting parameters. Although not perfect, we deem our current formalism as more conservative than the previous time-delay measurements of the COSMOGRAIL collaboration \citep{Bonvin2017, Courbin2017}. Whether this formalism is too conservative or not is still an open - and complicated - question, that we will address in Millon et al. (2019, in prep.).

In the end, we have one Group of time-delay estimates for each data set and curve-shifting technique. The next step is to combine these Groups together. As stated earlier, the two curve-shifting techniques implemented in \pycs\ are not fully independent, so we chose to fully marginalise over their respective results. In practice, this translates into summing the normalised probability density distributions of each time-delay estimates for the free-knot splines and the regression difference estimators Groups, thus yielding a single Group for each data set. Finally, combining the results of the various data sets together can be done either by multiplying the respective probability density distributions or by marginalising over the individual results, if we assume that the time-delay measurements are or are not independent, respectively.

\subsection{Application to \wfilens}
\label{sec:fiducialdelays}

We now apply this formalism to the data sets presented in Section \ref{sec:data}. For \wfilens, we have four different data sets. In a similar fashion to \citet{Bonvin2018}, we analyze these data sets independently from each other. However, we explore the results obtained by simultaneously fitting all of the data sets as a robustness test in Sec.~\ref{sec:robustness}.

The range of the curve-shifting technique parameters to explore must reflect the state of our knowledge about the data sets. For example, the light curves presented in Fig.~\ref{fig:lcs} are a combination of the intrinsic luminosity variations of the quasar, common to all light curves, and individual extrinsic variations due to microlensing magnification by compact objects in the lens galaxy. Since the amount of microlensing magnification is {\it a priori} unknown, various microlensing models should be considered if the curve-shifting technique aims to model it explicitly. Similarly, there is no clear consensus on how to properly represent the intrinsic quasar luminosity variations; on long temporal scales, a damped random walk model gives good results \citep[e.g.][]{Kozlowski2010, Zu2013}, whereas on short time scales, the power spectral density seems to be better fit by a power law \citep[e.g.][]{Mushotzky2011, Kasliwal2015}. Consequently, we adopt a data-driven approach and explore various choices for the parameters controlling the smoothness of the fit of the intrinsic and extrinsic variations.

The curve-shifting technique parameters used in this work are presented in Table \ref{tab:tabparams}. For each set of estimator parameters - 9 for the free-knot splines estimator and 5 for the regression difference estimator\footnote{We recall that the generative model parameters used are the same than the free-knot spline estimator, for simplicity.}, we obtain a Group of time-delay estimates, that we combine following the formalism presented in Section \ref{sec:pycs}. We present the combined time-delay measurements for each data set in Figure \ref{fig:delays}. We also present the two possible combinations of these data sets time-delay measurements: i) a marginalisation (labelled ``sum'' in the figure) and ii) a multiplication (labelled ``mult'' in the figure). In addition to our own measurements, we report the time-delay measurements of \citet{Vuissoz2008} and \citet{Morgan2018}, that use subsets of the data presented in this work. Both measurements are consistent with our work, although neither of them provide a direct measurement for the AC time delay; their AB and BC time delays are visible as smaller dots on Figure \ref{fig:delays}. We note that \citet{Giannini2017} also present light-curves of \wfilens\ taken with the 1.54m Danish telescope at the ESO La Silla Observatory, as part of the MiNDSTEp collaboration. However, they do not provide time-delay estimates. We tried to apply \pycs\ to their data but varying the estimator and generative model parameters gave discrepant results with very large uncertainties. We conclude that the MiNDSTEp data set is not of sufficient quality to produce reliable time-delay estimates on its own, and thus we do not include it in our analysis.

\begin{figure*}
  \centering
  \includegraphics[width=0.99\textwidth]{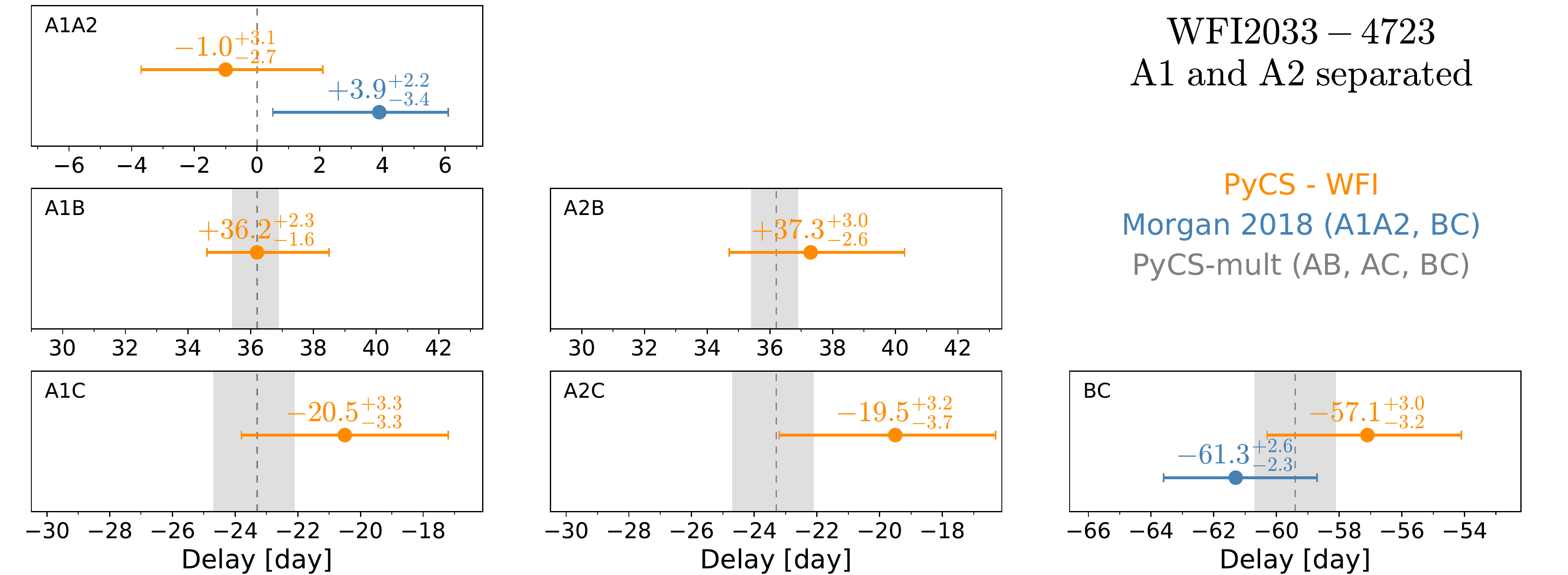}
  \caption{\wfilens\ time-delay estimates using the WFI data set with the A1 and A2 light curves separated. The A1A2 and BC measurements from \citet{Morgan2018} are reported for comparison. We also report the ``PyCS-mult'' estimates from Figure \ref{fig:delays}, obtained using the virtual A light curve, represented in this figure as shaded vertical bands. In this latter case, we report the AB and AC time-delay estimates in the A1B/A2B and A1C/A2C panels, respectively. We note that the BC estimate from WFI presented here slightly differs from Figure \ref{fig:delays} since the free-knot spline estimates used in the combined group are obtained by a joint fit to all of the light curves, and so are sensitive to the use of A1 and A2 instead of A. The values above each measurement give the $50^{\mathrm{th}}$, $16^{\mathrm{th}}$ and $84^{\mathrm{th}}$ percentiles of the respective probability distributions.}
  \label{fig:delays_a1a2}
\end{figure*}

Overall, our measurements agree well with each other. To quantify that agreement, we  compute the Bayes Factor, or evidence ratio $F = \mathcal{H}^{\mathrm{same}} / \mathcal{H}^{\mathrm{bias}}$, following \citet{Marshall2006}. The idea is to test which hypothesis is the most probable: either $\mathcal{H}^{\mathrm{same}}$ that the time-delay measurements of each data set are all representing the same quantity - the cosmological time delays - or $\mathcal{H}^{\mathrm{bias}}$ that at least one measurement represents a biased measurement of the cosmological time delay, for example due to the presence of microlensing time delay \citep{Tie2017}. $F>1$ indicates that it is more likely that the time-delay measurements are not significantly affected by systematic errors (or are similarly biased) than not. In such a case, a combination of the time-delays estimates such as the ``PyCS-mult'' should be favored over the marginalization of ``PyCS-sum''. We test every possible combination of the individual time-delay estimates and find that $F$ is always greater than 1. As it can be guessed from Figure \ref{fig:delays}, the smallest Bayes Factor arises when comparing the AC delay of SMARTS and WFI, with $F^{\mathrm{SMARTS+WFI}}_{\mathrm{AC}}=1.84$. The evidence ratios obtained when considering all the data sets independently are $F^{\mathrm{all}}_{\mathrm{AB}}\sim550$, $F^{\mathrm{all}}_{\mathrm{AC}}\sim76$ and $F^{\mathrm{all}}_{\mathrm{BC}}\sim123$, indicating that we should be able to use the combined ``PyCS-mult'' results of Figure \ref{fig:delays} without any loss of consistency.

We note that a Bayes Factor greater than one, showing that the individual measurements are mutually consistent, does not necessarily mean that there are no unaccounted random uncertainties affecting them. Rather, it means that the uncertainties of the individual measurement are large enough not to be significantly affected by unaccounted random error, or that similar systematic errors affect all our measurements. In the case of time-delay measurements, a potential albeit speculative source of error could be the microlensing time delay. A full description of it and an estimate of its amplitude for \wfilens\ is presented in Section \ref{sec:mltd}.

\subsection{A1 and A2 light curves of the WFI data set}

\citet{Morgan2018} measure a time delay between the A1 and A2 light curves of their SMARTS+EULER combined data set. To do so, they use a completely independent data reduction pipeline and curve-shifting techniques than the ones used in this work. Although we are also able to obtain reasonably well separated A1 and A2 light curves for the ECAM and SMARTS data set, applying our curve-shifting techniques on them results in a significant loss of precision over all measurable time delays with respect to our fiducial results obtained using the A light curve. However, the WFI data set is of better quality, and precise time-delay measurements can be obtained, as presented in Figure \ref{fig:delays_a1a2} along with the A1A2 and BC measurements reported in \citet{Morgan2018}. To ease comparison with Figure \ref{fig:delays}, we show the combined result ``PyCS-mult'' obtained with the A light curve as gray shaded vertical bands, with the AB and AC delay plotted in the A1B/A2B and A1C/A2C panels, respectively. 

For the WFI data set, we find a marginal evidence for the A2 light curve to lead A1, whereas \citet{Morgan2018} find the opposite. We note that basic lens mass models explored in \citet{Vuissoz2008} predict $1<\Delta t_{A1A2}<3$, i.e. A1 leading A2. The other delays are very close to the WFI ones obtained with A=A1+A2 - yet with worse precision and thus consistent with ``PyCS-mult'' results assuming a zero delay between A1 and A2. Despite the worse precision, using A1B and A2B (or A1C and A2C) in the lens models has the advantage over using AB (or AC) that it provides an extra constraint in the form of an independent time-delay estimate. It also solves the issue of the anchoring position of the A=A1+A2 virtual image: a time delay can now be allocated to each of the A1 and A2 images rather than at a mean position between the two, which could potentially bias the lens models.

\subsection{Robustness tests}
\label{sec:robustness}

\begin{figure*}
  \centering
  \includegraphics[width=0.99\textwidth]{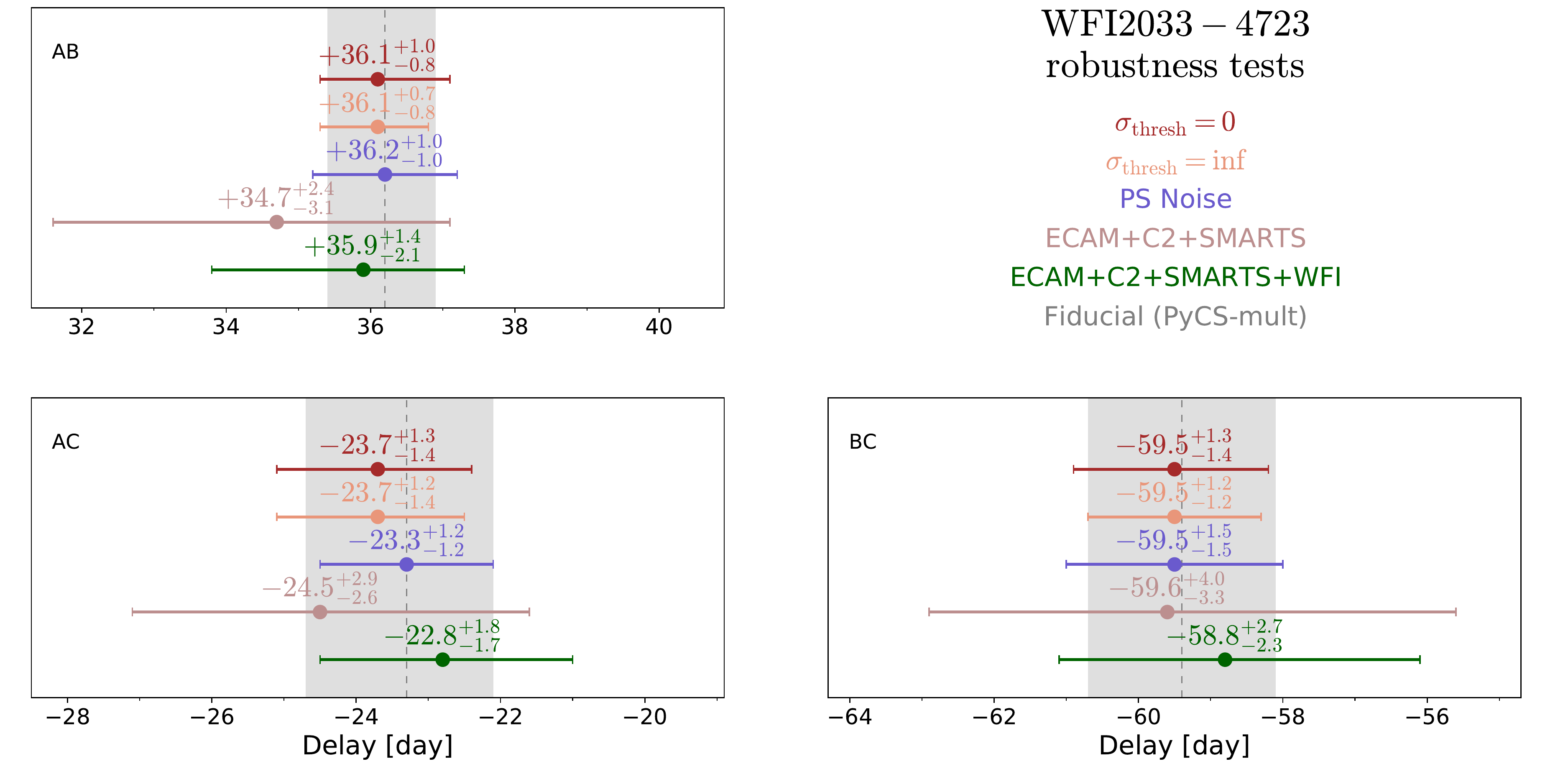}
  
   \vspace*{0.5cm}
  
  \includegraphics[width=0.99\textwidth]{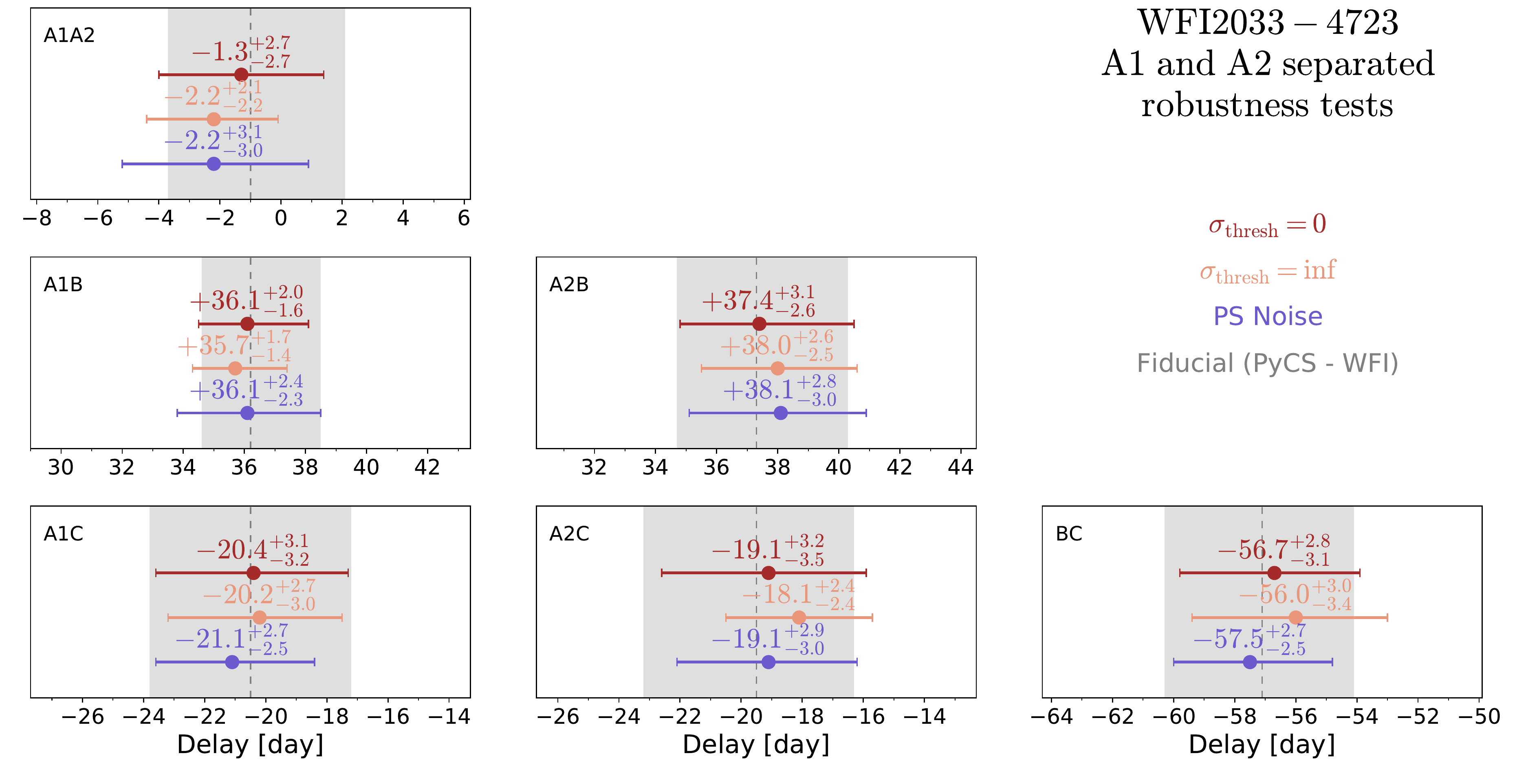}
  \caption{Results of the robustness tests for \wfilens\ time-delay measurements presented in Section \ref{sec:robustness}. The shaded vertical bands correspond to our fiducial results, labelled ``PyCS-mult'' and  ``PyCS - WFI'' in Figure \ref{fig:delays} and Figure \ref{fig:delays_a1a2} for the top and bottom groups of panels, respectively. The values indicated above each measurement represent the $50^{\mathrm{th}}$, $16^{\mathrm{th}}$ and $84^{\mathrm{th}}$ percentiles of the respective probability distributions.}
  \label{fig:robustness}
\end{figure*}

To complete our analysis of the time delays of \wfilens, we conduct a number of robustness tests to ensure the validity of our results (Section~\ref{sec:fiducialdelays}, hereafter the fiducial results).

The first test focuses on the Group combination threshold $\sigma_{\rm{thresh}}$ that was fixed at 0.5. Recall that this parameter controls how the Groups of time-delay estimates in a given Series (obtained by varying the estimator and generative model parameters of a given curve-shifting technique applied to a given data set) are combined. For $\sigma_{\rm{thresh}}=0$, all the Groups in the series are combined, as in a proper marginalisation. For $\sigma_{\rm{thresh}}\sim \infty$, only the most precise Group is used, in a similar behaviour to the previous COSMOGRAIL publications. These two cases represent the two ends of the spectrum of plausible combinations, none of which being optimal. On the one hand, marginalising over all the Groups in the Series includes Groups whose curve-shifting technique parameters are not necessarily well-suited to represent the data; on the other hand, considering only the most precise Group increases the risk of being affected by a systematic error linked to the specific choice of curve-shifting technique parameters. Ensuring that the results obtained in both cases are in agreement with our fiducial results is a good test of the consistency of the latter. The results obtained are presented in Figure \ref{fig:robustness}. They are in very good agreement with the fiducial results, considering both precision and accuracy. This also indicates that the different Groups in each Series were already in good agreement with each other prior to the marginalisation.

The second test focuses on the generative model, especially on the way the noise in the mock data is generated. Currently, as described in Sec.~\ref{sec:pycs}, it is a combination of white and red noise whose parameters are manually adjusted on a subset of mock light curves, until the statistical properties of the fit residuals match the ones computed on the real data \citep[see][for details]{Tewes2013a}. The noise parameters to adjust differ for each choice of data set and generative models parameters; in total, the present work required the adjustment of 36 sets of noise parameters. Although tedious, this procedure remains tractable when analysing a single lens system. However, it does not scale well with the analysis of a large collection of systems. For this reason, we developed a new, completely automated generative model that is based on the power spectrum of the data residuals to generate the noise in the mock curves. A detailed description of this new generative model will be presented in a forthcoming work (Millon et al. 2019, in prep.). The results of this new generative model are presented in Figure \ref{fig:robustness} and are labelled ``PS Noise''. They use the same estimator and generative model parameters as our fiducial models presented in Table \ref{tab:tabparams}, and are in excellent agreement with the fiducial results. This test assesses the robustness of our fiducial noise generation scheme, but also highlights the performance of the new scheme. We thus decide to use the new scheme for the remaining robustness tests presented in this section since it is much easier to apply than the fiducial one.

Finally, for the analysis of the previous H0LiCOW lenses we used a different framework to measure time delays in which all the data sets were combined into a single set of light curves. Although in the current framework we have good reasons to separately analyse the data sets, we want to assess the effect of merging the data sets on the final time-delay estimates. This provides an important insight on the robustness of the previous time-delay measurements used by the H0LiCOW collaboration (see \citet{Tewes2013b} for \rxjlens\ and \citet{Bonvin2017} for \hequad\ - an updated measurement of the time delays of these two systems will be presented in Millon et al. 2019, in prep.). We merge the data sets by applying an offset in magnitude that is computed by minimizing the scatter between overlapping parts of each light curve and each data set. We explore two combinations, one with the four data sets combined and one without the WFI data set; since the latter stands out the most in terms of signal-to-noise ratio, sampling and duration, we want to assess its weight in the final combination. The results are presented in the upper panel of Figure \ref{fig:robustness}. We can see that the addition of the WFI data set has a clear impact, shifting the combined results towards the delays obtained on the WFI data set alone and significantly improving the overall precision. It is however less precise than our fiducial results. 

\section{Microlensing time delay}
\label{sec:mltd}

In this section, we compute the impact of the microlensing time delay, a speculative effect first described in \citet{Tie2017}. The analysis performed in this section is similar to the one presented in \citet{Bonvin2018} for the lensed quasar PG1115+080, itself based on the original work of \citet{Tie2017}. In the following, we use the terminology introduced in \citet{Bonvin2019}.

If it were possible to observe the lensed images with a resolution of a few microarcseconds, the spatial profile of the source would reveal the accretion disk of the quasar. One possible model characterisation of the accretion disk emission is the lamp-post model \citep[e.g.][]{Cackett2007, Starkey2016}, that predicts a temperature change propagating across the disk driven by temperature variations at the center, thus triggering radiation. As a result, the emissions from the outer regions of the disk follow the emission from the inner regions, but with a delay. As the disk is seen as a point source at the spatial resolution we are working with, the observed light curves are a blend of the light emitted from all the spatial regions of the source visible through the chosen filter. Due to the delayed emission in the outer regions, the variations of the observed, blended light curves are seen slightly delayed with respect to the variations of hypothetical, unblended light curves of the disk's inner regions. This excess of time delay associated to the lamp-post model of variability is called \emph{geometrical time delay}, as it relates to the spatial extension of the source. Since the geometrical delay is the same in all lensed images, it cancels out when measuring the delay between two images, and has no impact on the cosmological time delay measurement, which is the delay of interest for time-delay cosmography.

However, stars and/or other compact objects in the lens galaxy act as microlenses and differentially magnify the various regions of the lensed accretion disk. The net effect of this micromagnification is to reweight the geometrical delay for each lensed image; the additional excess of time delay introduced by the reweighting is called the \emph{microlensing time delay}. Since the position of the microlenses in the lens galaxy is random, the excesses of microlensing time delays are uncorrelated and so do not cancel out between images. This results in a bias when measuring cosmological time delays between two lensed images.

In order to quantify the contribution of the microlensing time delay on the measured time delays, we first simulate microlensing magnification maps at the image positions, using the lens model from H0LiCOW XII (see Table B1). The convergence, shear and stellar mass fraction used to generate the microlensing maps are given in Table~\ref{tab:mlparams}.

\begin{table}
 \caption{The $\kappa$, $\gamma$, and $\kappa_{\star}/\kappa$ at each lensed image position from the macro model presented in H0LiCOW XII.}
  \centering
 \begin{tabular}{l c c c}
  \hline
  Image & $\kappa$ & $\gamma$ & $\kappa_{\star}/\kappa$ \\
  \hline
 A1 & 0.350 & 0.340 & 0.612 \\ 
 A2 & 0.462 & 0.424 & 0.690 \\ 
 B & 0.281 & 0.309 & 0.519 \\ 
 C & 0.567 & 0.547 & 0.698 \\ 
  \hline
 \end{tabular}
 \label{tab:mlparams}
\end{table}

\begin{figure*}
\centering
\includegraphics[width=0.99\textwidth]{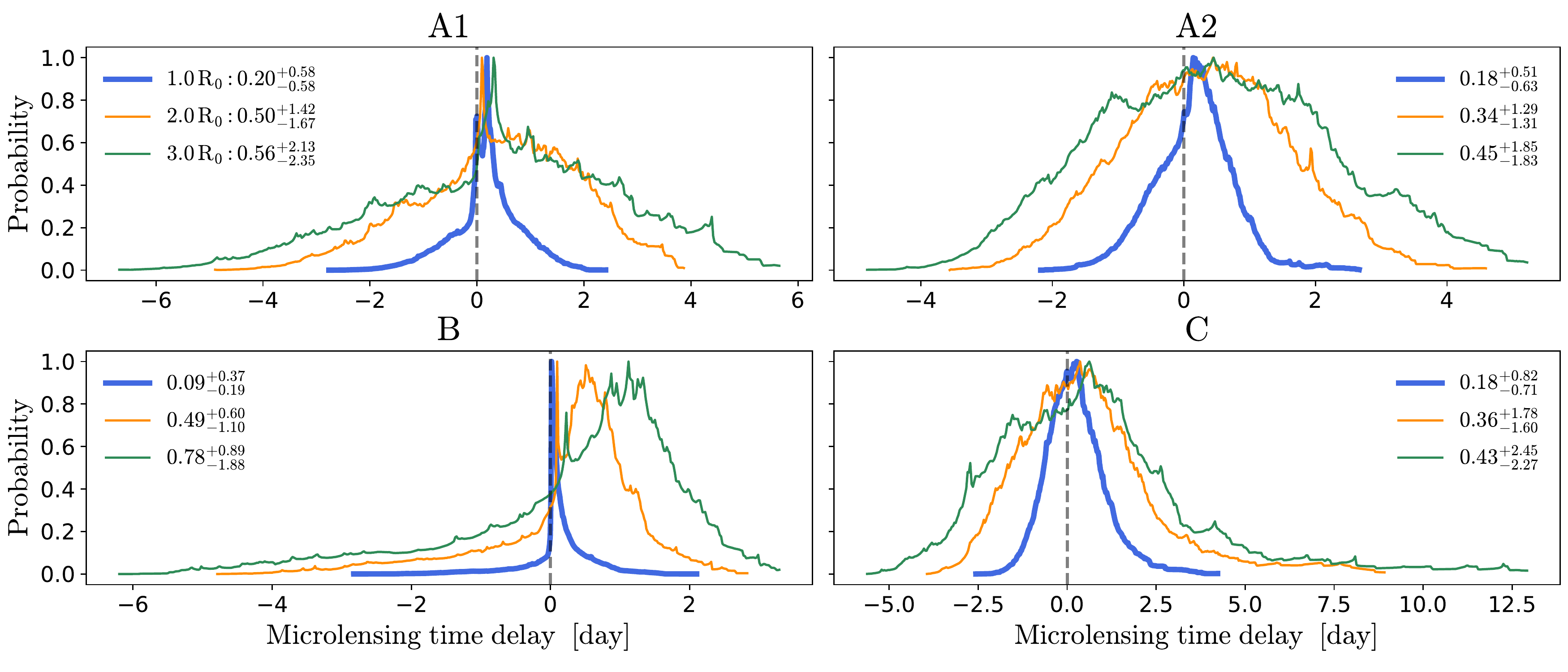}

\vspace{0.5cm}

\begin{tabular}{r c c c c c c c c c c c}

$R_0$ & symbol & A & A1A2 & A1B & A1C & A2B & A2C & AB & AC & BC \\ \hline

\vspace{0.1cm} 1.0 & \raisebox{1.0pt}{\textcolor{royalblue}{\scaleobj{1.6}{\rule{0.3cm}{0.8pt}}}} & $0.18^{+0.40}_{-0.43}$ & $-0.01^{+0.71}_{-0.75}$ & $-0.02^{+0.55}_{-0.58}$ & $0.03^{+0.99}_{-0.84}$ & $-0.04^{+0.73}_{-0.67}$ & $0.05^{+1.03}_{-0.93}$ & $0.13^{+0.55}_{-0.49}$ & $0.21^{+0.92}_{-0.81}$ & $0.07^{+0.96}_{-0.82}$ \\

\vspace{0.1cm} 2.0 & \raisebox{1.0pt}{\textcolor{darkorange}{\scaleobj{1.6}{\rule{0.3cm}{0.8pt}}}} & $0.41^{+0.96}_{-1.03}$ & $-0.17^{+1.63}_{-1.48}$ & $-0.11^{+1.04}_{-1.19}$ & $-0.12^{+2.26}_{-1.75}$ & $0.01^{+1.55}_{-1.71}$ & $0.06^{+2.31}_{-2.05}$ & $0.54^{+0.98}_{-1.19}$ & $0.51^{+2.00}_{-1.71}$ & $0.04^{+2.30}_{-1.81}$ \\

\vspace{0.1cm} 3.0 & \raisebox{1.0pt}{\textcolor{seagreen}{\scaleobj{1.6}{\rule{0.3cm}{0.8pt}}}} & $0.50^{+1.41}_{-1.43}$ & $-0.32^{+2.35}_{-2.06}$ & $-0.21^{+1.53}_{-1.87}$ & $-0.35^{+3.09}_{-2.47}$ & $0.03^{+2.22}_{-2.56}$ & $-0.06^{+3.15}_{-2.88}$ & $0.81^{+1.37}_{-1.92}$ & $0.57^{+2.67}_{-2.38}$ & $-0.08^{+3.27}_{-2.60}$ \\

\end{tabular}

\caption{Distributions of the excess of microlensing time delay for the four lensed image of \wfilens. The values displayed represent the 16th, 50th and 84th percentiles of the plotted distributions. The thicker lines (in blue) indicate the fiducial case where the source size corresponds to the thin-disk model prediction. The vertical dashed grey lines represent the case where no microlensing time-delay is present. The table below the figure reports the percentiles of the distributions of the combined image A as well as the lensed image pair.}
\label{fig:mltd_distrib}
\end{figure*}

We use GPU-D \citep{Vernardos2014}, a GPU-accelerated implementation of the inverse ray-shooting technique from \mbox{\citet{Wambsganss1992}}\footnote{Technically speaking, \citet{Wambsganss1992} use a tree code to simplify the computations whereas \citep{Vernardos2014} do the full computation.} that produces a magnification map in the source plane from a spatially random distribution of microlenses in the lens plane. We assume that the microlenses follow Salpeter mass function with a ratio of the upper to lower masses of $r=100$ and a mean microlens mass of $\langle M \rangle = 0.3 M_{\odot}$. Each map has $8192\times8192$ pixels, representing a physical size of $20$ Einstein radii $\langle R_{Ein} \rangle$, defined as:

\begin{equation}
\langle R_{\rm Ein} \rangle =\sqrt{\frac{D_{\rm s}D_{\rm ds}}{D_{\rm d}}\frac{4G\langle M \rangle }{c^2}} = 2.375\times 10^{16} {\rm cm},
\end{equation}

\noindent where $D_{\rm d}, D_{\rm s}$ and $D_{\rm ds}$ are the angular diameter distances\footnote{Angular diameter distances directly depend on \hc. We use a value of \hc=72 \ksmpc but note that changing this value only marginally affects the microlensing time-delay distributions.} between the observer and the deflector (i.e. the lens galaxy), the observer and the source and the deflector and the source, respectively. A $20 \langle R_{\mathrm{Ein}} \rangle$ square region is sufficiently large to statistically sample the effects of microlensing. We do not present in this work what these magnification maps look like, but the interested reader can have a look at Figure 2 of \citet{Bonvin2019} for an example.

The quasar accretion disk is modeled using the thin-disk model of \citet{Shakura1973}. The thin-disk model requires i) the wavelength at which the observations are made, that we take as the center of the WFI {\tt BB\#Rc/162} filter (6517.25 \AA)\footnote{choosing the {\tt KPNO R-band} or {\tt Rouge Gen\`eve} filter instead does not produce a significant difference.}, ii) an Eddington luminosity ratio $L/L_E$ fixed at 0.3, iii) a radiative efficiency for the black hole at the center of the accretion disk of $\eta=0.1$ and iv) a black hole mass of $M_{BH} = 4.26\times10^8 M_{\odot}$ taken from \citet{Sluse2012}. This gives an estimated radius of $R_0=1.291\times 10^{15}$ cm. Our choices of $L/L_E$, $\eta$ and $M_{BH}$ follow \citet{Morgan2018}, but that other values for these parameters are possible. For example, \citet{Motta2017} estimate a black-hole mass of $M_{BH} = 1.2\times10^8 M_{\odot}$, albeit with a much larger uncertainty, and based on microlensing estimates of the accretion disk size. Adopting this value will greatly reduce the predicted size of the accretion disk, and consequently the predicted microlensing time delay as well by a factor of $\sim2.3$. We however decide to stick to the \citet{Sluse2012} black-hole mass estimate since it is based on a more standard virial estimation of the mass using observations of MgII emission lines.

Next, we associate the lamp-post model of variability to the thin-disk model and place the disk at a specific position on the magnification maps. For each region of the disk, one can compute the excess of microlensing time delay by reweighting the excess of geometrical time delay at this region. Integrating over all the pixels of the disk, we obtain the excess of microlensing time delay \citep[see Equation 10 in][]{Tie2017}. By marginalizing over all the possible source positions in the microlensing maps, we obtain probability distributions of the excess of microlensing time delay for each lensed images. These are presented in Figure \ref{fig:mltd_distrib}. We also explore two extra configurations of the thin-disk model where we multiply the theoretical characteristic radius $R_0$ by a factor of two and three. This is motivated by the fact that direct estimates of disk sizes using either microlensing \citep[e.g.][]{Rojas2014, JimenezVicente2016, Morgan2018} or also reverberation mapping \citep[e.g.][]{Edelson2015, Lira2015, Fausnaugh2016} generally infer sizes 2-3 times larger than the thin-disk theory predictions. The black-hole mass versus source size relation fitted on the analysis of the modeled microlensing of 14 lensed quasars in \citet{Morgan2018} predicts a source size $\sim$three times larger than the thin-disk model prediction, which corresponds to the higher value we explore here. We keep the disc inclination and position angle at zero across all our tests, as varying these produces only a second-order effect on the amplitude of the excess of microlensing time delay \citep[e.g][]{Tie2017, Bonvin2018}.


A proper inclusion of the microlensing time delay into the cosmological analysis should be done cautiously. The distributions and values presented in Fig.~\ref{fig:mltd_distrib} are computed assuming a source position fixed in the microlensing map. However, due to the transverse motion of the source behind the lens galaxy, microlensing time delay changes over time. As highlighted in \citet{Bonvin2019}, what drives the effect of microlensing time delay is not the absolute amount of micromagnification, but the spatial variability of the micromagnification across the source profile. Thus, the time it takes for the source to fully cross a critical curve in the magnification map is a characteristics of the period over which one can expect microlensing time delay to potentially strongly affect the time delay measurements \citep[see e.g.][for estimates of typical time-scales]{Mosquera2011}.

A conservative approach to include microlensing time delay in our measurements is to simply convolve the probability distributions presented in the Table of Figure \ref{fig:mltd_distrib} with the time-delay measurement uncertainties presented in Section \ref{sec:fiducialdelays}. A finer approach has been proposed in \citet{Chen2018}, that takes into account the discrepancies between the various measured time-delay and their predicted counterparts from lens modeling. However, there is so far no formalism that takes into account the duration of the monitoring campaign.  In the case of \wfilens, we follow the approach from \citet{Chen2018}. This analysis is presented in H0LiCOW XII, where the results with a source size of $1R_0$ have been included by default. H0LiCOW XII also shows that including or not microlensing time-delay has a minimal impact on the time-delay distance inference of \wfilens. As a conclusion, we note that our measured A1-A2 delay from the WFI data set is consistent with zero, as predicted by the macro lens model; although not incompatible with the existence of microlensing time delay, it does not favor large source sizes.

\section{Conclusions}
\label{sec:conclusions}

This work is part of a series carrying out a new analysis of quadruply lensed system \wfilens\ for time-delay cosmography. H0LiCOW XI (Sluse et al., submitted) studies the environment of the lens galaxy and H0LiCOW XII (Rusu et al., submitted) models the lens system. This work focuses on the measurement of the time delays between the multiple images of the lensed source. The key points summarize as follows:

\begin{itemize}
 \item We present the most complete collection of monitoring data of \wfilens\ to date. It totals $\sim450$ hours of monitoring, split in four data sets (WFI, ECAM, C2 and SMARTS) from four different instruments, spanning more than 14 years of observations. It notably includes a full season of high-cadence (daily) and high-precision (milimagnitude) monitoring with the WFI instrument mounted on the MPIA 2.2m telescope at La Silla Observatory. The monitoring data are turned into light curves using the pipeline built around the MCS deconvolution algorithm \citep{Magain1998, Cantale2016}, allowing us to accurately deblend the light from the lensed quasar images even in poor seeing conditions.
 
 \item The time delays are estimated individually on each data set using the open-source \pycs\ software \citep{Tewes2013a, Bonvin2016}. The measurements on each data set are consistent with each other. We choose to combine them into a single group of time-delay estimates, resulting in $\Delta t_{AB} = 36.2_{-0.8}^{+0.7}$ (2.1\% precision),  $\Delta t_{AC} = -23.3_{-1.4}^{+1.2}$ (5.6\%) and $\Delta t_{BC} = -59.4_{-1.3}^{+1.3}$ (2.2\%) days. 
 
 \item The higher signal-to-noise ratio and better seeing of the WFI exposures allows to disentangle the A1 and A2 images of the lensed quasar into two light curves. The time-delay measurement between the two images yields $\Delta t_{A1A2} = -1.0_{-2.7}^{+3.1}$ days, consistent with a null delay.
 
 \item Unlike the time-delay estimates used in the past by H0LiCOW, the estimates presented in this work are obtained by marginalizing over the various parameters of our curve-shifting techniques following the formalism introduced in \citet{Bonvin2018}. In addition, a series of tests are conducted to ensure the robustness of our results. Updating the time-delay estimates of the other H0LiCOW lens systems is planned for a future milestone update.
 
 \item We estimate the contribution of the microlensing time delay to the measured delays using the quasar black-hole mass estimate from \citet{Sluse2012}. The predictions obtained using the standard thin-disk model are moderate, and are implemented by default at the lens modeling stage (H0LiCOW XII). Increasing the disk size to match the measurements from microlensing studies \citep{Morgan2018} increases the contribution of microlensing time delay. The precision of the measured time delays does not allow neither to confirm nor to rule out the presence of microlensing time delay in the data. 
 
\end{itemize}

The measured time delays and estimated microlensing time delays are used at the lens modeling stage (H0LiCOW XII) to measure the time-delay distance of \wfilens\ and infer a value of the Hubble constant \hc. H0LiCOW XIII (Wong et al. 2019, in prep.) combines the time-delay distance of \wfilens\ with two other lenses systems fully analysed by H0LiCOW (\blens\ from \citet{Suyu2010} and SDSS~1206+4332 from \citet{Birrer2019}) with a combined analysis of HST+AO images of \rxjlens, \hequad\ and PG1115+080 from a joint STIDES+H0LiCOW effort (Chen et al. 2019, in prep.), turning them into the most precise joint inference of \hc from time-delay cosmography to date. It also updates the joint inference with other cosmological probes, superseding the previous results of H0LiCOW V \citep{Bonvin2017}.

This paper is the third of a series reporting successful COSMOGRAIL monitoring campaigns with the MPIA 2.2m telescope; it follows the time-delay measurements of DES~J0408-5354 \citep{Courbin2017} and PG1115+080 \citep{Bonvin2018}. Significantly improving the precision with which the Hubble constant can be determined from time-delay cosmography will require the analysis of dozens of lensed systems. Being able to measure robust time-delay estimates in a short amount of time is crucial in that regard. Our goal is to measure several tens of new delays in the next 5 years. Prospects are excellent as the precision on time delays in individual objects is of the order of 2-5\% in only 1 season of high-cadence (daily) and high SNR ($>$500 per quasar image) monitoring for each object. With smaller telescopes (1m) and lower cadence (1 point every 4 days) such a precision required typically 10 years of effort per object.

\begin{acknowledgements}
COSMOGRAIL and H0LiCOW are made possible thanks to the continuous work of all observers and technical staff obtaining the monitoring observations, in particular at the Swiss Euler telescope at La Silla Observatory. We also warmly thank R. Gredel, H.-W. Rix and T. Henning for allowing us to observe with the MPIA 2.2m telescope and for ensuring the daily cadence of the observations is achieved. This research is support by the Swiss National Science Foundation (SNSF) and by the European Research Council (ERC) under the European Union’s Horizon 2020 research and innovation programme (COSMICLENS: grant agreement No 787866). SHS and DCYC thank the Max Planck Society for support through the Max Planck Research Group for SHS. This work was supported by World Premier International Research Center Initiative (WPI Initiative), MEXT, Japan. K.C.W. is supported in part by an EACOA Fellowship awarded by the East Asia Core Observatories Association, which consists of the Academia Sinica Institute of Astronomy and Astrophysics, the National Astronomical Observatory of Japan, the National Astronomical Observatories of the Chinese Academy of Sciences, and the Korea Astronomy and Space Science Institute. T. A. acknowledges support by the Ministry for the Economy, Development, and Tourism's Programa Inicativa Cient\'{i}fica Milenio through grant IC 12009, awarded to The Millennium Institute of Astrophysics (MAS). C.S.K is supported by NSF grants AST-1515876, AST-1515927 and AST-181440 as well as a fellowship from the Radcliffe Institute for Advanced Study at Harvard University. C.M is supported by the NSF grant $\#1614018$.

This research made use of Astropy, a community-developed core Python package for Astronomy \citep{Astropy2013, Astropy2018}, the 2D graphics environment Matplotlib \citep{Hunter2007} and the source extraction software Sextractor \citep{Bertin1996}.

\end{acknowledgements}

\bibliographystyle{aa}
\bibliography{wfi2033delays_aa}

\begin{thebibliography}{71}
\expandafter\ifx\csname natexlab\endcsname\relax\def\natexlab#1{#1}\fi

\bibitem[{{Abbott} {et~al.}(2017){Abbott}, {Abbott}, {Abbott}, {Acernese},
  {Ackley}, {Adams}, {Adams}, {Addesso}, {Adhikari}, {Adya}, \&
  et~al.}]{Abbott2017}
{Abbott}, B.~P., {Abbott}, R., {Abbott}, T.~D., {et~al.} 2017, \nat, 551, 85

\bibitem[{{Amendola} {et~al.}(2019){Amendola}, {Dirian}, {Nersisyan}, \&
  {Park}}]{Amendola2019}
{Amendola}, L., {Dirian}, Y., {Nersisyan}, H., \& {Park}, S. 2019, arXiv
  e-prints

\bibitem[{{Astropy Collaboration} {et~al.}(2018){Astropy Collaboration},
  {Price-Whelan}, {Sip{\H o}cz}, {G{\"u}nther}, {Lim}, {Crawford}, {Conseil},
  {Shupe}, {Craig}, {Dencheva}, {Ginsburg}, {VanderPlas}, {Bradley},
  {P{\'e}rez-Su{\'a}rez}, {de Val-Borro}, {Aldcroft}, {Cruz}, {Robitaille},
  {Tollerud}, {Ardelean}, {Babej}, {Bachetti}, {Bakanov}, {Bamford},
  {Barentsen}, {Barmby}, {Baumbach}, {Berry}, {Biscani}, {Boquien}, {Bostroem},
  {Bouma}, {Brammer}, {Bray}, {Breytenbach}, {Buddelmeijer}, {Burke},
  {Calderone}, {Cano Rodr{\'{\i}}guez}, {Cara}, {Cardoso}, {Cheedella},
  {Copin}, {Crichton}, {D{\'A}vella}, {Deil}, {Depagne}, {Dietrich}, {Donath},
  {Droettboom}, {Earl}, {Erben}, {Fabbro}, {Ferreira}, {Finethy}, {Fox},
  {Garrison}, {Gibbons}, {Goldstein}, {Gommers}, {Greco}, {Greenfield},
  {Groener}, {Grollier}, {Hagen}, {Hirst}, {Homeier}, {Horton}, {Hosseinzadeh},
  {Hu}, {Hunkeler}, {Ivezi{\'c}}, {Jain}, {Jenness}, {Kanarek}, {Kendrew},
  {Kern}, {Kerzendorf}, {Khvalko}, {King}, {Kirkby}, {Kulkarni}, {Kumar},
  {Lee}, {Lenz}, {Littlefair}, {Ma}, {Macleod}, {Mastropietro}, {McCully},
  {Montagnac}, {Morris}, {Mueller}, {Mumford}, {Muna}, {Murphy}, {Nelson},
  {Nguyen}, {Ninan}, {N{\"o}the}, {Ogaz}, {Oh}, {Parejko}, {Parley}, {Pascual},
  {Patil}, {Patil}, {Plunkett}, {Prochaska}, {Rastogi}, {Reddy Janga},
  {Sabater}, {Sakurikar}, {Seifert}, {Sherbert}, {Sherwood-Taylor}, {Shih},
  {Sick}, {Silbiger}, {Singanamalla}, {Singer}, {Sladen}, {Sooley},
  {Sornarajah}, {Streicher}, {Teuben}, {Thomas}, {Tremblay}, {Turner},
  {Terr{\'o}n}, {van Kerkwijk}, {de la Vega}, {Watkins}, {Weaver}, {Whitmore},
  {Woillez}, \& {Zabalza}}]{Astropy2018}
{Astropy Collaboration}, {Price-Whelan}, A.~M., {Sip{\H o}cz}, B.~M., {et~al.}
  2018, ArXiv e-prints:1801.02634

\bibitem[{{Astropy Collaboration} {et~al.}(2013){Astropy Collaboration},
  {Robitaille}, {Tollerud}, {Greenfield}, {Droettboom}, {Bray}, {Aldcroft},
  {Davis}, {Ginsburg}, {Price-Whelan}, {Kerzendorf}, {Conley}, {Crighton},
  {Barbary}, {Muna}, {Ferguson}, {Grollier}, {Parikh}, {Nair}, {Unther},
  {Deil}, {Woillez}, {Conseil}, {Kramer}, {Turner}, {Singer}, {Fox}, {Weaver},
  {Zabalza}, {Edwards}, {Azalee Bostroem}, {Burke}, {Casey}, {Crawford},
  {Dencheva}, {Ely}, {Jenness}, {Labrie}, {Lim}, {Pierfederici}, {Pontzen},
  {Ptak}, {Refsdal}, {Servillat}, \& {Streicher}}]{Astropy2013}
{Astropy Collaboration}, {Robitaille}, T.~P., {Tollerud}, E.~J., {et~al.} 2013,
  \aap, 558, A33

\bibitem[{{Bautista} {et~al.}(2017){Bautista}, {Busca}, {Guy}, {Rich},
  {Blomqvist}, {du Mas des Bourboux}, {Pieri}, {Font-Ribera}, {Bailey},
  {Delubac}, {Kirkby}, {Le Goff}, {Margala}, {Slosar}, {Vazquez}, {Brownstein},
  {Dawson}, {Eisenstein}, {Miralda-Escud{\'e}}, {Noterdaeme},
  {Palanque-Delabrouille}, {P{\^a}ris}, {Petitjean}, {Ross}, {Schneider},
  {Weinberg}, \& {Y{\`e}che}}]{Bautista2017}
{Bautista}, J.~E., {Busca}, N.~G., {Guy}, J., {et~al.} 2017, \aap, 603, A12

\bibitem[{{Bertin} \& {Arnouts}(1996)}]{Bertin1996}
{Bertin}, E. \& {Arnouts}, S. 1996, \aaps, 117, 393

\bibitem[{{Birrer} {et~al.}(2019){Birrer}, {Treu}, {Rusu}, {Bonvin},
  {Fassnacht}, {Chan}, {Agnello}, {Shajib}, {Chen}, {Auger}, {Courbin},
  {Hilbert}, {Sluse}, {Suyu}, {Wong}, {Marshall}, {Lemaux}, \&
  {Meylan}}]{Birrer2019}
{Birrer}, S., {Treu}, T., {Rusu}, C.~E., {et~al.} 2019, \mnras, 484, 4726

\bibitem[{{Bonvin} {et~al.}(2018){Bonvin}, {Chan}, {Millon}, {Rojas},
  {Courbin}, {Chen}, {Fassnacht}, {Paic}, {Tewes}, {Chao}, {Chijani}, {Gilman},
  {Gilmore}, {Williams}, {Buckley-Geer}, {Frieman}, {Marshall}, {Suyu}, {Treu},
  {Hempel}, {Kim}, {Lachaume}, {Rabus}, {Anguita}, {Meylan}, {Motta}, \&
  {Magain}}]{Bonvin2018}
{Bonvin}, V., {Chan}, J.~H.~H., {Millon}, M., {et~al.} 2018, \aap, 616, A183

\bibitem[{{Bonvin} {et~al.}(2017){Bonvin}, {Courbin}, {Suyu}, {Marshall},
  {Rusu}, {Sluse}, {Tewes}, {Wong}, {Collett}, {Fassnacht}, {Treu}, {Auger},
  {Hilbert}, {Koopmans}, {Meylan}, {Rumbaugh}, {Sonnenfeld}, \&
  {Spiniello}}]{Bonvin2017}
{Bonvin}, V., {Courbin}, F., {Suyu}, S.~H., {et~al.} 2017, \mnras, 465, 4914

\bibitem[{{Bonvin} {et~al.}(2016){Bonvin}, {Tewes}, {Courbin}, {Kuntzer},
  {Sluse}, \& {Meylan}}]{Bonvin2016}
{Bonvin}, V., {Tewes}, M., {Courbin}, F., {et~al.} 2016, \aap, 585, A88

\bibitem[{{Bonvin} {et~al.}(2019){Bonvin}, {Tihhonova}, {Millon}, {Chan},
  {Savary}, {Huber}, \& {Courbin}}]{Bonvin2019}
{Bonvin}, V., {Tihhonova}, O., {Millon}, M., {et~al.} 2019, \aap, 621, A55

\bibitem[{{Braatz} {et~al.}(2018){Braatz}, {Pesce}, {Condon}, \&
  {Reid}}]{Braatz2018}
{Braatz}, J., {Pesce}, D., {Condon}, J., \& {Reid}, M. 2018, in Astronomical
  Society of the Pacific Conference Series, Vol. 517, Science with a Next
  Generation Very Large Array, ed. E.~{Murphy}, 821

\bibitem[{{Cackett} {et~al.}(2007){Cackett}, {Horne}, \&
  {Winkler}}]{Cackett2007}
{Cackett}, E.~M., {Horne}, K., \& {Winkler}, H. 2007, \mnras, 380, 669

\bibitem[{{Cantale} {et~al.}(2016){Cantale}, {Courbin}, {Tewes}, {Jablonka.},
  \& {Meylan}}]{Cantale2016}
{Cantale}, N., {Courbin}, F., {Tewes}, M., {Jablonka.}, P., \& {Meylan}, G.
  2016, ArXiv1602.02167

\bibitem[{{Cao} {et~al.}(2017){Cao}, {Zheng}, {Biesiada}, {Qi}, {Chen}, \&
  {Zhu}}]{Cao2017}
{Cao}, S., {Zheng}, X., {Biesiada}, M., {et~al.} 2017, \aap, 606, A15

\bibitem[{{Capozziello} \& {Ruchika}(2019)}]{Capozziello2019}
{Capozziello}, S. \& {Ruchika}, Anjan~A., S. 2019, \mnras, 484, 4484

\bibitem[{{Chen} {et~al.}(2018{\natexlab{a}}){Chen}, {Chan}, {Bonvin},
  {Fassnacht}, {Rojas}, {Millon}, {Courbin}, {Suyu}, {Wong}, {Sluse}, {Treu},
  {Shajib}, {Hsueh}, {Lagattuta}, {Koopmans}, {Vegetti}, \&
  {McKean}}]{Chen2018}
{Chen}, G.~C.-F., {Chan}, J.~H.~H., {Bonvin}, V., {et~al.} 2018{\natexlab{a}},
  \mnras, 481, 1115

\bibitem[{{Chen} {et~al.}(2018{\natexlab{b}}){Chen}, {Fishbach}, \&
  {Holz}}]{ChenHY2018}
{Chen}, H.-Y., {Fishbach}, M., \& {Holz}, D.~E. 2018{\natexlab{b}}, \nat, 562,
  545

\bibitem[{{Courbin} {et~al.}(2018){Courbin}, {Bonvin}, {Buckley-Geer},
  {Fassnacht}, {Frieman}, {Lin}, {Marshall}, {Suyu}, {Treu}, {Anguita},
  {Motta}, {Meylan}, {Paic}, {Tewes}, {Agnello}, {Chao}, {Chijani}, {Gilman},
  {Rojas}, {Williams}, {Hempel}, {Kim}, {Lachaume}, {Rabus}, {Abbott}, {Allam},
  {Annis}, {Banerji}, {Bechtol}, {Benoit-L{\'e}vy}, {Brooks}, {Burke}, {Carnero
  Rosell}, {Carrasco Kind}, {Carretero}, {D'Andrea}, {da Costa}, {Davis},
  {DePoy}, {Desai}, {Flaugher}, {Fosalba}, {Garc{\'{\i}}a-Bellido},
  {Gaztanaga}, {Goldstein}, {Gruen}, {Gruendl}, {Gschwend}, {Gutierrez},
  {Honscheid}, {James}, {Kuehn}, {Kuhlmann}, {Kuropatkin}, {Lahav}, {Lima},
  {Maia}, {March}, {Marshall}, {McMahon}, {Menanteau}, {Miquel}, {Nord},
  {Plazas}, {Sanchez}, {Scarpine}, {Schindler}, {Schubnell}, {Sevilla-Noarbe},
  {Smith}, {Soares-Santos}, {Sobreira}, {Suchyta}, {Tarle}, {Tucker}, {Walker},
  \& {Wester}}]{Courbin2017}
{Courbin}, F., {Bonvin}, V., {Buckley-Geer}, E., {et~al.} 2018, \aap, 609, A71

\bibitem[{{DES Collaboration} {et~al.}(2018){DES Collaboration}, {Abbott},
  {Abdalla}, {Annis}, {Bechtol}, {Blazek}, {Benson}, {Bernstein}, {Bernstein},
  {Bertin}, {Brooks}, {Burke}, {Carnero Rosell}, {Carrasco Kind}, {Carretero},
  {Castander}, {Chang}, {Crawford}, {Cunha}, {D'Andrea}, {da Costa}, {Davis},
  {DeRose}, {Desai}, {Diehl}, {Dietrich}, {Doel}, {Drlica-Wagner}, {Evrard},
  {Fernandez}, {Flaugher}, {Fosalba}, {Frieman}, {Garc{\'{\i}}a-Bellido},
  {Gaztanaga}, {Gerdes}, {Giannantonio}, {Gruen}, {Gruendl}, {Gschwend},
  {Gutierrez}, {Hartley}, {Henning}, {Honscheid}, {Hoyle}, {Huterer}, {Jain},
  {James}, {Jarvis}, {Jeltema}, {Johnson}, {Johnson}, {Krause}, {Kuehn},
  {Kuhlmann}, {Kuropatkin}, {Lahav}, {Liddle}, {Lima}, {Lin}, {MacCrann},
  {Maia}, {Manzotti}, {March}, {Marshall}, {Miquel}, {Mohr}, {Natoli},
  {Nugent}, {Ogando}, {Park}, {Plazas}, {Reichardt}, {Reil}, {Roodman}, {Ross},
  {Rozo}, {Rykoff}, {Sanchez}, {Scarpine}, {Schubnell}, {Scolnic},
  {Sevilla-Noarbe}, {Sheldon}, {Smith}, {Smith}, {Soares-Santos}, {Sobreira},
  {Suchyta}, {Tarle}, {Thomas}, {Troxel}, {Walker}, {Wechsler}, {Weller},
  {Wester}, {Wu}, \& {Zuntz}}]{DES2018}
{DES Collaboration}, {Abbott}, T.~M.~C., {Abdalla}, F.~B., {et~al.} 2018,
  \mnras

\bibitem[{{Dhawan} {et~al.}(2018){Dhawan}, {Jha}, \& {Leibundgut}}]{Dhawan2018}
{Dhawan}, S., {Jha}, S.~W., \& {Leibundgut}, B. 2018, \aap, 609, A72

\bibitem[{{Edelson} {et~al.}(2015){Edelson}, {Gelbord}, {Horne}, {McHardy},
  {Peterson}, {Ar{\'e}valo}, {Breeveld}, {De Rosa}, {Evans}, {Goad}, {Kriss},
  {Brandt}, {Gehrels}, {Grupe}, {Kennea}, {Kochanek}, {Nousek}, {Papadakis},
  {Siegel}, {Starkey}, {Uttley}, {Vaughan}, {Young}, {Barth}, {Bentz},
  {Brewer}, {Crenshaw}, {Dalla Bont{\`a}}, {De Lorenzo-C{\'a}ceres}, {Denney},
  {Dietrich}, {Ely}, {Fausnaugh}, {Grier}, {Hall}, {Kaastra}, {Kelly},
  {Korista}, {Lira}, {Mathur}, {Netzer}, {Pancoast}, {Pei}, {Pogge},
  {Schimoia}, {Treu}, {Vestergaard}, {Villforth}, {Yan}, \& {Zu}}]{Edelson2015}
{Edelson}, R., {Gelbord}, J.~M., {Horne}, K., {et~al.} 2015, \apj, 806, 129

\bibitem[{{Fausnaugh} {et~al.}(2016){Fausnaugh}, {Denney}, {Barth}, {Bentz},
  {Bottorff}, {Carini}, {Croxall}, {De Rosa}, {Goad}, {Horne}, {Joner},
  {Kaspi}, {Kim}, {Klimanov}, {Kochanek}, {Leonard}, {Netzer}, {Peterson},
  {Schn{\"u}lle}, {Sergeev}, {Vestergaard}, {Zheng}, {Zu}, {Anderson},
  {Ar{\'e}valo}, {Bazhaw}, {Borman}, {Boroson}, {Brandt}, {Breeveld}, {Brewer},
  {Cackett}, {Crenshaw}, {Dalla Bont{\`a}}, {De Lorenzo-C{\'a}ceres},
  {Dietrich}, {Edelson}, {Efimova}, {Ely}, {Evans}, {Filippenko}, {Flatland},
  {Gehrels}, {Geier}, {Gelbord}, {Gonzalez}, {Gorjian}, {Grier}, {Grupe},
  {Hall}, {Hicks}, {Horenstein}, {Hutchison}, {Im}, {Jensen}, {Jones},
  {Kaastra}, {Kelly}, {Kennea}, {Kim}, {Korista}, {Kriss}, {Lee}, {Lira},
  {MacInnis}, {Manne-Nicholas}, {Mathur}, {McHardy}, {Montouri}, {Musso},
  {Nazarov}, {Norris}, {Nousek}, {Okhmat}, {Pancoast}, {Papadakis}, {Parks},
  {Pei}, {Pogge}, {Pott}, {Rafter}, {Rix}, {Saylor}, {Schimoia}, {Siegel},
  {Spencer}, {Starkey}, {Sung}, {Teems}, {Treu}, {Turner}, {Uttley},
  {Villforth}, {Weiss}, {Woo}, {Yan}, \& {Young}}]{Fausnaugh2016}
{Fausnaugh}, M.~M., {Denney}, K.~D., {Barth}, A.~J., {et~al.} 2016, \apj, 821,
  56

\bibitem[{{Feeney} {et~al.}(2019){Feeney}, {Peiris}, {Williamson}, {Nissanke},
  {Mortlock}, {Alsing}, \& {Scolnic}}]{Feeney2019}
{Feeney}, S.~M., {Peiris}, H.~V., {Williamson}, A.~R., {et~al.} 2019, Physical
  Review Letters, 122, 061105

\bibitem[{{Giannini} {et~al.}(2017){Giannini}, {Schmidt}, {Wambsganss},
  {Alsubai}, {Andersen}, {Anguita}, {Bozza}, {Bramich}, {Browne}, {Calchi
  Novati}, {Damerdji}, {Diehl}, {Dodds}, {Dominik}, {Elyiv}, {Fang}, {Figuera
  Jaimes}, {Finet}, {Gerner}, {Gu}, {Hardis}, {Harps{\o}e}, {Hinse},
  {Hornstrup}, {Hundertmark}, {Jessen-Hansen}, {J{\o}rgensen}, {Juncher},
  {Kains}, {Kerins}, {Korhonen}, {Liebig}, {Lund}, {Lundkvist}, {Maier},
  {Mancini}, {Masi}, {Mathiasen}, {Penny}, {Proft}, {Rabus}, {Rahvar}, {Ricci},
  {Scarpetta}, {Sahu}, {Sch{\"a}fer}, {Sch{\"o}nebeck}, {Skottfelt},
  {Snodgrass}, {Southworth}, {Surdej}, {Tregloan-Reed}, {Vilela}, {Wertz}, \&
  {Zimmer}}]{Giannini2017}
{Giannini}, E., {Schmidt}, R.~W., {Wambsganss}, J., {et~al.} 2017, \aap, 597,
  A49

\bibitem[{{Hikage} {et~al.}(2019){Hikage}, {Oguri}, {Hamana}, {More},
  {Mandelbaum}, {Takada}, {K{\"o}hlinger}, {Miyatake}, {Nishizawa}, {Aihara},
  {Armstrong}, {Bosch}, {Coupon}, {Ducout}, {Ho}, {Hsieh}, {Komiyama},
  {Lanusse}, {Leauthaud}, {Lupton}, {Medezinski}, {Mineo}, {Miyama},
  {Miyazaki}, {Murata}, {Murayama}, {Shirasaki}, {Sif{\'o}n}, {Simet},
  {Speagle}, {Spergel}, {Strauss}, {Sugiyama}, {Tanaka}, {Utsumi}, {Wang}, \&
  {Yamada}}]{Hikage2019}
{Hikage}, C., {Oguri}, M., {Hamana}, T., {et~al.} 2019, \pasj, 71, 43

\bibitem[{Hunter(2007)}]{Hunter2007}
Hunter, J.~D. 2007, Computing In Science \& Engineering, 9, 90

\bibitem[{{Jang} {et~al.}(2018){Jang}, {Hatt}, {Beaton}, {Lee}, {Freedman},
  {Madore}, {Hoyt}, {Monson}, {Rich}, {Scowcroft}, \& {Seibert}}]{Jang2018}
{Jang}, I.~S., {Hatt}, D., {Beaton}, R.~L., {et~al.} 2018, \apj, 852, 60

\bibitem[{{Jim{\'e}nez-Vicente} {et~al.}(2015){Jim{\'e}nez-Vicente},
  {Mediavilla}, {Kochanek}, \& {Mu{\~n}oz}}]{JimenezVicente2016}
{Jim{\'e}nez-Vicente}, J., {Mediavilla}, E., {Kochanek}, C.~S., \& {Mu{\~n}oz},
  J.~A. 2015, \apj, 806, 251

\bibitem[{{Kasliwal} {et~al.}(2015){Kasliwal}, {Vogeley}, \&
  {Richards}}]{Kasliwal2015}
{Kasliwal}, V.~P., {Vogeley}, M.~S., \& {Richards}, G.~T. 2015, \mnras, 451,
  4328

\bibitem[{{K{\"o}hlinger} {et~al.}(2017){K{\"o}hlinger}, {Viola}, {Joachimi},
  {Hoekstra}, {van Uitert}, {Hildebrandt}, {Choi}, {Erben}, {Heymans},
  {Joudaki}, {Klaes}, {Kuijken}, {Merten}, {Miller}, {Schneider}, \&
  {Valentijn}}]{Kohlinger2017}
{K{\"o}hlinger}, F., {Viola}, M., {Joachimi}, B., {et~al.} 2017, \mnras, 471,
  4412

\bibitem[{{Koz{\l}owski} {et~al.}(2010){Koz{\l}owski}, {Kochanek}, {Udalski},
  {Wyrzykowski}, {Soszy{\'n}ski}, {Szyma{\'n}ski}, {Kubiak}, {Pietrzy{\'n}ski},
  {Szewczyk}, {Ulaczyk}, {Poleski}, \& {OGLE Collaboration}}]{Kozlowski2010}
{Koz{\l}owski}, S., {Kochanek}, C.~S., {Udalski}, A., {et~al.} 2010, \apj, 708,
  927

\bibitem[{{Kozmanyan} {et~al.}(2019){Kozmanyan}, {Bourdin}, {Mazzotta},
  {Rasia}, \& {Sereno}}]{Kozmanyan2019}
{Kozmanyan}, A., {Bourdin}, H., {Mazzotta}, P., {Rasia}, E., \& {Sereno}, M.
  2019, \aap, 621, A34

\bibitem[{{Lira} {et~al.}(2015){Lira}, {Ar{\'e}valo}, {Uttley}, {McHardy}, \&
  {Videla}}]{Lira2015}
{Lira}, P., {Ar{\'e}valo}, P., {Uttley}, P., {McHardy}, I.~M.~M., \& {Videla},
  L. 2015, \mnras, 454, 368

\bibitem[{{Magain} {et~al.}(1998){Magain}, {Courbin}, \& {Sohy}}]{Magain1998}
{Magain}, P., {Courbin}, F., \& {Sohy}, S. 1998, \apj, 494, 472

\bibitem[{{Marshall} {et~al.}(2006){Marshall}, {Rajguru}, \&
  {Slosar}}]{Marshall2006}
{Marshall}, P., {Rajguru}, N., \& {Slosar}, A. 2006, \prd, 73, 067302

\bibitem[{Molinari {et~al.}(2004)Molinari, Durand, \& Sabatier}]{Molinari2004}
Molinari, N., Durand, J., \& Sabatier, R. 2004, Computational Statistics {\&}
  Data Analysis, 45, 159

\bibitem[{{Morgan} {et~al.}(2018){Morgan}, {Hyer}, {Bonvin}, {Mosquera},
  {Cornachione}, {Courbin}, {Kochanek}, \& {Falco}}]{Morgan2018}
{Morgan}, C.~W., {Hyer}, G.~E., {Bonvin}, V., {et~al.} 2018, \apj, 869, 106

\bibitem[{{Morgan} {et~al.}(2004){Morgan}, {Caldwell}, {Schechter}, {Dressler},
  {Egami}, \& {Rix}}]{Morgan2004}
{Morgan}, N.~D., {Caldwell}, J.~A.~R., {Schechter}, P.~L., {et~al.} 2004, \aj,
  127, 2617

\bibitem[{{M{\"o}rtsell} \& {Dhawan}(2018)}]{Mortsell2018}
{M{\"o}rtsell}, E. \& {Dhawan}, S. 2018, \jcap, 9, 025

\bibitem[{{Mosquera} \& {Kochanek}(2011)}]{Mosquera2011}
{Mosquera}, A.~M. \& {Kochanek}, C.~S. 2011, \apj, 738, 96

\bibitem[{{Motta} {et~al.}(2017){Motta}, {Mediavilla}, {Rojas}, {Falco},
  {Jim{\'e}nez-Vicente}, \& {Mu{\~n}oz}}]{Motta2017}
{Motta}, V., {Mediavilla}, E., {Rojas}, K., {et~al.} 2017, \apj, 835, 132

\bibitem[{{Mushotzky} {et~al.}(2011){Mushotzky}, {Edelson}, {Baumgartner}, \&
  {Gandhi}}]{Mushotzky2011}
{Mushotzky}, R.~F., {Edelson}, R., {Baumgartner}, W., \& {Gandhi}, P. 2011,
  \apjl, 743, L12

\bibitem[{{Pandey} {et~al.}(2019){Pandey}, {Karwal}, \& {Das}}]{Pandey2019}
{Pandey}, K.~L., {Karwal}, T., \& {Das}, S. 2019, arXiv e-prints

\bibitem[{{Planck Collaboration} {et~al.}(2018{\natexlab{a}}){Planck
  Collaboration}, {Aghanim}, {Akrami}, {Ashdown}, {Aumont}, {Baccigalupi},
  {Ballardini}, {Banday}, {Barreiro}, {Bartolo}, {Basak}, {Battye}, {Benabed},
  {Bernard}, {Bersanelli}, {Bielewicz}, {Bock}, {Bond}, {Borrill}, {Bouchet},
  {Boulanger}, {Bucher}, {Burigana}, {Butler}, {Calabrese}, {Cardoso},
  {Carron}, {Challinor}, {Chiang}, {Chluba}, {Colombo}, {Combet}, {Contreras},
  {Crill}, {Cuttaia}, {de Bernardis}, {de Zotti}, {Delabrouille}, {Delouis},
  {Di Valentino}, {Diego}, {Dor{\'e}}, {Douspis}, {Ducout}, {Dupac}, {Dusini},
  {Efstathiou}, {Elsner}, {En{\ss}lin}, {Eriksen}, {Fantaye}, {Farhang},
  {Fergusson}, {Fernandez-Cobos}, {Finelli}, {Forastieri}, {Frailis},
  {Franceschi}, {Frolov}, {Galeotta}, {Galli}, {Ganga}, {G{\'e}nova-Santos},
  {Gerbino}, {Ghosh}, {Gonz{\'a}lez-Nuevo}, {G{\'o}rski}, {Gratton},
  {Gruppuso}, {Gudmundsson}, {Hamann}, {Handley}, {Herranz}, {Hivon}, {Huang},
  {Jaffe}, {Jones}, {Karakci}, {Keih{\"a}nen}, {Keskitalo}, {Kiiveri}, {Kim},
  {Kisner}, {Knox}, {Krachmalnicoff}, {Kunz}, {Kurki-Suonio}, {Lagache},
  {Lamarre}, {Lasenby}, {Lattanzi}, {Lawrence}, {Le Jeune}, {Lemos},
  {Lesgourgues}, {Levrier}, {Lewis}, {Liguori}, {Lilje}, {Lilley}, {Lindholm},
  {L{\'o}pez-Caniego}, {Lubin}, {Ma}, {Mac{\'{\i}}as-P{\'e}rez}, {Maggio},
  {Maino}, {Mandolesi}, {Mangilli}, {Marcos-Caballero}, {Maris}, {Martin},
  {Martinelli}, {Mart{\'{\i}}nez-Gonz{\'a}lez}, {Matarrese}, {Mauri}, {McEwen},
  {Meinhold}, {Melchiorri}, {Mennella}, {Migliaccio}, {Millea}, {Mitra},
  {Miville-Desch{\^e}nes}, {Molinari}, {Montier}, {Morgante}, {Moss}, {Natoli},
  {N{\o}rgaard-Nielsen}, {Pagano}, {Paoletti}, {Partridge}, {Patanchon},
  {Peiris}, {Perrotta}, {Pettorino}, {Piacentini}, {Polastri}, {Polenta},
  {Puget}, {Rachen}, {Reinecke}, {Remazeilles}, {Renzi}, {Rocha}, {Rosset},
  {Roudier}, {Rubi{\~n}o-Mart{\'{\i}}n}, {Ruiz-Granados}, {Salvati}, {Sandri},
  {Savelainen}, {Scott}, {Shellard}, {Sirignano}, {Sirri}, {Spencer},
  {Sunyaev}, {Suur-Uski}, {Tauber}, {Tavagnacco}, {Tenti}, {Toffolatti},
  {Tomasi}, {Trombetti}, {Valenziano}, {Valiviita}, {Van Tent}, {Vibert},
  {Vielva}, {Villa}, {Vittorio}, {Wandelt}, {Wehus}, {White}, {White},
  {Zacchei}, \& {Zonca}}]{Planck2018cosmo}
{Planck Collaboration}, {Aghanim}, N., {Akrami}, Y., {et~al.}
  2018{\natexlab{a}}, ArXiv e-prints

\bibitem[{{Planck Collaboration} {et~al.}(2018{\natexlab{b}}){Planck
  Collaboration}, {Akrami}, {Arroja}, {Ashdown}, {Aumont}, {Baccigalupi},
  {Ballardini}, {Banday}, {Barreiro}, {Bartolo}, {Basak}, {Battye}, {Benabed},
  {Bernard}, {Bersanelli}, {Bielewicz}, {Bock}, {Bond}, {Borrill}, {Bouchet},
  {Boulanger}, {Bucher}, {Burigana}, {Butler}, {Calabrese}, {Cardoso},
  {Carron}, {Casaponsa}, {Challinor}, {Chiang}, {Colombo}, {Combet},
  {Contreras}, {Crill}, {Cuttaia}, {de Bernardis}, {de Zotti}, {Delabrouille},
  {Delouis}, {D{\'e}sert}, {Di Valentino}, {Dickinson}, {Diego}, {Donzelli},
  {Dor{\'e}}, {Douspis}, {Ducout}, {Dupac}, {Efstathiou}, {Elsner},
  {En{\ss}lin}, {Eriksen}, {Falgarone}, {Fantaye}, {Fergusson},
  {Fernandez-Cobos}, {Finelli}, {Forastieri}, {Frailis}, {Franceschi},
  {Frolov}, {Galeotta}, {Galli}, {Ganga}, {G{\'e}nova-Santos}, {Gerbino},
  {Ghosh}, {Gonz{\'a}lez-Nuevo}, {G{\'o}rski}, {Gratton}, {Gruppuso},
  {Gudmundsson}, {Hamann}, {Handley}, {Hansen}, {Helou}, {Herranz}, {Hivon},
  {Huang}, {Jaffe}, {Jones}, {Karakci}, {Keih{\"a}nen}, {Keskitalo}, {Kiiveri},
  {Kim}, {Kisner}, {Knox}, {Krachmalnicoff}, {Kunz}, {Kurki-Suonio}, {Lagache},
  {Lamarre}, {Langer}, {Lasenby}, {Lattanzi}, {Lawrence}, {Le Jeune}, {Leahy},
  {Lesgourgues}, {Levrier}, {Lewis}, {Liguori}, {Lilje}, {Lilley}, {Lindholm},
  {L{\'o}pez-Caniego}, {Lubin}, {Ma}, {Mac{\'{\i}}as-P{\'e}rez}, {Maggio},
  {Maino}, {Mandolesi}, {Mangilli}, {Marcos-Caballero}, {Maris}, {Martin},
  {Mart{\'{\i}}nez-Gonz{\'a}lez}, {Matarrese}, {Mauri}, {McEwen}, {Meerburg},
  {Meinhold}, {Melchiorri}, {Mennella}, {Migliaccio}, {Millea}, {Mitra},
  {Miville-Desch{\^e}nes}, {Molinari}, {Moneti}, {Montier}, {Morgante}, {Moss},
  {Mottet}, {M{\"u}nchmeyer}, {Natoli}, {N{\o}rgaard-Nielsen}, {Oxborrow},
  {Pagano}, {Paoletti}, {Partridge}, {Patanchon}, {Pearson}, {Peel}, {Peiris},
  {Perrotta}, {Pettorino}, {Piacentini}, {Polastri}, {Polenta}, {Puget},
  {Rachen}, {Reinecke}, {Remazeilles}, {Renzi}, {Rocha}, {Rosset}, {Roudier},
  {Rubi{\~n}o-Mart{\'{\i}}n}, {Ruiz-Granados}, {Salvati}, {Sandri},
  {Savelainen}, {Scott}, {Shellard}, {Shiraishi}, {Sirignano}, {Sirri},
  {Spencer}, {Sunyaev}, {Suur-Uski}, {Tauber}, {Tavagnacco}, {Tenti},
  {Terenzi}, {Toffolatti}, {Tomasi}, {Trombetti}, {Valiviita}, {Van Tent},
  {Vibert}, {Vielva}, {Villa}, {Vittorio}, {Wandelt}, {Wehus}, {White},
  {White}, {Zacchei}, \& {Zonca}}]{Planck2018overview}
{Planck Collaboration}, {Akrami}, Y., {Arroja}, F., {et~al.}
  2018{\natexlab{b}}, ArXiv e-prints

\bibitem[{{Poulin} {et~al.}(2018){Poulin}, {Smith}, {Karwal}, \&
  {Kamionkowski}}]{Poulin2018}
{Poulin}, V., {Smith}, T.~L., {Karwal}, T., \& {Kamionkowski}, M. 2018, arXiv
  e-prints

\bibitem[{{Refsdal}(1964)}]{Refsdal1964}
{Refsdal}, S. 1964, \mnras, 128, 307

\bibitem[{{Reid} {et~al.}(2013){Reid}, {Braatz}, {Condon}, {Lo}, {Kuo},
  {Impellizzeri}, \& {Henkel}}]{Reid2013}
{Reid}, M.~J., {Braatz}, J.~A., {Condon}, J.~J., {et~al.} 2013, \apj, 767, 154

\bibitem[{{Riess} {et~al.}(2019){Riess}, {Casertano}, {Yuan}, {Macri}, \&
  {Scolnic}}]{Riess2019}
{Riess}, A.~G., {Casertano}, S., {Yuan}, W., {Macri}, L.~M., \& {Scolnic}, D.
  2019, arXiv e-prints

\bibitem[{{Rojas} {et~al.}(2014){Rojas}, {Motta}, {Mediavilla}, {Falco},
  {Jim{\'e}nez-Vicente}, \& {Mu{\~n}oz}}]{Rojas2014}
{Rojas}, K., {Motta}, V., {Mediavilla}, E., {et~al.} 2014, \apj, 797, 61

\bibitem[{{Rusu} {et~al.}(2017){Rusu}, {Fassnacht}, {Sluse}, {Hilbert}, {Wong},
  {Huang}, {Suyu}, {Collett}, {Marshall}, {Treu}, \& {Koopmans}}]{Rusu2017}
{Rusu}, C.~E., {Fassnacht}, C.~D., {Sluse}, D., {et~al.} 2017, \mnras, 467,
  4220

\bibitem[{{Shakura} \& {Sunyaev}(1973)}]{Shakura1973}
{Shakura}, N.~I. \& {Sunyaev}, R.~A. 1973, \aap, 24, 337

\bibitem[{{Sluse} {et~al.}(2012){Sluse}, {Hutsem{\'e}kers}, {Courbin},
  {Meylan}, \& {Wambsganss}}]{Sluse2012}
{Sluse}, D., {Hutsem{\'e}kers}, D., {Courbin}, F., {Meylan}, G., \&
  {Wambsganss}, J. 2012, \aap, 544, A62

\bibitem[{{Sluse} {et~al.}(2017){Sluse}, {Sonnenfeld}, {Rumbaugh}, {Rusu},
  {Fassnacht}, {Treu}, {Suyu}, {Wong}, {Auger}, {Bonvin}, {Collett}, {Courbin},
  {Hilbert}, {Koopmans}, {Marshall}, {Meylan}, {Spiniello}, \&
  {Tewes}}]{Sluse2017}
{Sluse}, D., {Sonnenfeld}, A., {Rumbaugh}, N., {et~al.} 2017, \mnras, 470, 4838

\bibitem[{{Starkey} {et~al.}(2016){Starkey}, {Horne}, \&
  {Villforth}}]{Starkey2016}
{Starkey}, D.~A., {Horne}, K., \& {Villforth}, C. 2016, \mnras, 456, 1960

\bibitem[{{Suyu} {et~al.}(2017){Suyu}, {Bonvin}, {Courbin}, {Fassnacht},
  {Rusu}, {Sluse}, {Treu}, {Wong}, {Auger}, {Ding}, {Hilbert}, {Marshall},
  {Rumbaugh}, {Sonnenfeld}, {Tewes}, {Tihhonova}, {Agnello}, {Blandford},
  {Chen}, {Collett}, {Koopmans}, {Liao}, {Meylan}, \& {Spiniello}}]{Suyu2017}
{Suyu}, S.~H., {Bonvin}, V., {Courbin}, F., {et~al.} 2017, \mnras, 468, 2590

\bibitem[{{Suyu} {et~al.}(2018){Suyu}, {Chang}, {Courbin}, \&
  {Okumura}}]{Suyu2018}
{Suyu}, S.~H., {Chang}, T.-C., {Courbin}, F., \& {Okumura}, T. 2018, \ssr, 214,
  91

\bibitem[{{Suyu} {et~al.}(2010){Suyu}, {Marshall}, {Auger}, {Hilbert},
  {Blandford}, {Koopmans}, {Fassnacht}, \& {Treu}}]{Suyu2010}
{Suyu}, S.~H., {Marshall}, P.~J., {Auger}, M.~W., {et~al.} 2010, \apj, 711, 201

\bibitem[{{Suyu} {et~al.}(2014){Suyu}, {Treu}, {Hilbert}, {Sonnenfeld},
  {Auger}, {Blandford}, {Collett}, {Courbin}, {Fassnacht}, {Koopmans},
  {Marshall}, {Meylan}, {Spiniello}, \& {Tewes}}]{Suyu2014}
{Suyu}, S.~H., {Treu}, T., {Hilbert}, S., {et~al.} 2014, \apjl, 788, L35

\bibitem[{{Tewes} {et~al.}(2013{\natexlab{a}}){Tewes}, {Courbin}, \&
  {Meylan}}]{Tewes2013a}
{Tewes}, M., {Courbin}, F., \& {Meylan}, G. 2013{\natexlab{a}}, \aap, 553, A120

\bibitem[{{Tewes} {et~al.}(2013{\natexlab{b}}){Tewes}, {Courbin}, {Meylan},
  {Kochanek}, {Eulaers}, {Cantale}, {Mosquera}, {Magain}, {Van Winckel},
  {Sluse}, {Cataldi}, {V{\"o}r{\"o}s}, \& {Dye}}]{Tewes2013b}
{Tewes}, M., {Courbin}, F., {Meylan}, G., {et~al.} 2013{\natexlab{b}}, \aap,
  556, A22

\bibitem[{{Tie} \& {Kochanek}(2018)}]{Tie2017}
{Tie}, S.~S. \& {Kochanek}, C.~S. 2018, \mnras, 473, 80

\bibitem[{{Tihhonova} {et~al.}(2017){Tihhonova}, {Courbin}, {Harvey},
  {Hilbert}, {Rusu}, {Fassnacht}, {Bonvin}, {Marshall}, {Meylan}, {Sluse},
  {Suyu}, {Treu}, \& {Wong}}]{Tihhonova2018}
{Tihhonova}, O., {Courbin}, F., {Harvey}, D., {et~al.} 2017, ArXiv e-prints

\bibitem[{{Treu} \& {Marshall}(2016)}]{Treu2016}
{Treu}, T. \& {Marshall}, P.~J. 2016, \aapr, 24, 11

\bibitem[{{Troxel} {et~al.}(2018){Troxel}, {Krause}, {Chang}, {Eifler},
  {Friedrich}, {Gruen}, {MacCrann}, {Chen}, {Davis}, {DeRose}, {Dodelson},
  {Gatti}, {Hoyle}, {Huterer}, {Jarvis}, {Lacasa}, {Lemos}, {Peiris}, {Prat},
  {Samuroff}, {S{\'a}nchez}, {Sheldon}, {Vielzeuf}, {Wang}, {Zuntz}, {Lahav},
  {Abdalla}, {Allam}, {Annis}, {Avila}, {Bertin}, {Brooks}, {Burke}, {Carnero
  Rosell}, {Carrasco Kind}, {Carretero}, {Crocce}, {Cunha}, {D'Andrea}, {da
  Costa}, {De Vicente}, {Diehl}, {Doel}, {Evrard}, {Flaugher}, {Fosalba},
  {Frieman}, {Garc{\'{\i}}a-Bellido}, {Gaztanaga}, {Gerdes}, {Gruendl},
  {Gschwend}, {Gutierrez}, {Hartley}, {Hollowood}, {Honscheid}, {James},
  {Kirk}, {Kuehn}, {Kuropatkin}, {Li}, {Lima}, {March}, {Menanteau}, {Miquel},
  {Mohr}, {Ogando}, {Plazas}, {Roodman}, {Sanchez}, {Scarpine}, {Schindler},
  {Sevilla-Noarbe}, {Smith}, {Soares-Santos}, {Sobreira}, {Suchyta}, {Swanson},
  {Thomas}, {Walker}, \& {Wechsler}}]{Troxel2018}
{Troxel}, M.~A., {Krause}, E., {Chang}, C., {et~al.} 2018, \mnras, 479, 4998

\bibitem[{{Vernardos} {et~al.}(2014){Vernardos}, {Fluke}, {Bate}, \&
  {Croton}}]{Vernardos2014}
{Vernardos}, G., {Fluke}, C.~J., {Bate}, N.~F., \& {Croton}, D. 2014, \apjs,
  211, 16

\bibitem[{{Vuissoz} {et~al.}(2008){Vuissoz}, {Courbin}, {Sluse}, {Meylan},
  {Chantry}, {Eulaers}, {Morgan}, {Eyler}, {Kochanek}, {Coles}, {Saha},
  {Magain}, \& {Falco}}]{Vuissoz2008}
{Vuissoz}, C., {Courbin}, F., {Sluse}, D., {et~al.} 2008, \aap, 488, 481

\bibitem[{{Wambsganss} {et~al.}(1992){Wambsganss}, {Witt}, \&
  {Schneider}}]{Wambsganss1992}
{Wambsganss}, J., {Witt}, H.~J., \& {Schneider}, P. 1992, \aap, 258, 591

\bibitem[{{Wong} {et~al.}(2017){Wong}, {Suyu}, {Auger}, {Bonvin}, {Courbin},
  {Fassnacht}, {Halkola}, {Rusu}, {Sluse}, {Sonnenfeld}, {Treu}, {Collett},
  {Hilbert}, {Koopmans}, {Marshall}, \& {Rumbaugh}}]{Wong2017}
{Wong}, K.~C., {Suyu}, S.~H., {Auger}, M.~W., {et~al.} 2017, \mnras, 465, 4895

\bibitem[{Ying {et~al.}(2013)Ying, Kochanek, Kozlowski, \& Udalski}]{Zu2013}
Ying, Z., Kochanek, C.~S., Kozlowski, S., \& Udalski, Andrzej, E.-m. y.-s.
  2013, Astrophysical Journal, 765

\end{thebibliography}

\end{document}